\documentclass{aa}  
\usepackage{graphicx}
\usepackage{subfig}
\usepackage[export]{adjustbox}
\usepackage{txfonts}
\begin{document}

   \title{Reconstructing the star formation rate for compact binary populations with the Einstein telescope}
     \titlerunning{Reconstructing star formation rate with ET}

   \author{Neha Singh,
          \inst{1}
          \thanks{\email{singh@lapth.cnrs.fr}}
          Tomasz Bulik,
          \inst{2}
          Krzysztof Belczynski,
          \inst{3}
          Marek Cieslar,
          \inst{3}
           Francesca Calore
          \inst{1}
          }
   \institute{Laboratoire d’Annecy-le-Vieux de Physique Théorique (LAPTh), USMB, CNRS, F-74940 Annecy, France. \and
             Astronomical Observatory, University of Warsaw, Al. Ujazdowskie 4, 00-478 Warsaw, Poland. \and 
             Nicolaus Copernicus Astronomical Center, Polish Academy of Sciences, ul. Bartycka 18, 00-716 Warsaw, Poland.
             }
    \authorrunning {N. Singh et al. }

   \date{Received ; accepted  }

 
\abstract
{The Einstein Telescope (ET) is a proposed third-generation, wide-band gravitational wave (GW) detector. Given its improved detection sensitivity in comparison to the second-generation detectors, it will be capable of exploring the Universe with GWs up to very high redshifts. In this paper, we present a population-independent method to infer the functional form of star formation rate density (SFR) for different populations of compact binaries originating in stars from Population (Pop) I+II and Pop III using ET as a single instrument. We use an algorithm to answer three major questions regarding the SFR of different populations of compact binaries. Specifically, these questions refer to the termination redshift of the formation of Pop III stars, the redshift at peak SFR, and the functional form of SFR at high redshift, all of which remain to be elucidated.  We show that the reconstruction of SFR as a function of redshift for the different populations of compact binaries is independent of the time-delay distributions up to $z \sim 14,$ and that the accuracy of the reconstruction only strongly depends on this distribution at higher redshifts of $z\gtrsim 14$. We define the termination redshift for Pop III stars  as the redshift where the SFR drops to 1\% of its peak value. In this analysis, we constrain the peak of the SFR as a function of redshift and show that ET as a single instrument can distinguish the termination redshifts of different SFRs for Pop III stars, which have a true separation of at least $\Delta z \sim 2$. The accurate estimation of the termination redshift depends on correctly modelling the tail of the time-delay distribution, which constitutes delay times of $\gtrsim 8$ Gyr.
}

\maketitle
%

\section{Introduction}

Third-generation detectors of wide-band gravitational
waves (GWs) such as the Einstein Telescope (ET; \cite{2011CQGra..28i4013H, 2010CQGra..27s4002P}) or the Cosmic Explorer (CE; \cite{PhysRevD.91.082001, Abbott2017, 2019BAAS...51g..35R}) will be able to probe a much larger volume of the Universe, and will therefore have a much higher detection rate \citep{2020JCAP...03..050M} compared to the current second-generation detectors. ET will have a detection sensitivity down to 1Hz \citep{2008arXiv0810.0604H,2012CQGra..29l4006H} and will therefore have the ability to detect binary black holes (BBHs) of high mass; that is, in the range of  $10^2-10^4 M_{\odot}$ \citep{2011PhRvD..83d4020H, 2011PhRvD..83d4021H, 2011GReGr..43..485G, 2010ApJ...722.1197A}. Assuming ET-D \citep{2011CQGra..28i4013H} design sensitivity, the expected detection rates are $\sim 10^5 - 10^6$ BBH detections and $\sim 7 \times 10^4$ binary neutron star (BNS) detections in one year \citep{2012PhRvD..86l2001R, 2014PhRvD..89h4046R, 2019JCAP...08..015B}. Given its increased detection sensitivity, the ET will also be able to detect the coalescence of compact binaries with a total mass of 20 - 100 $M_{\odot}$, typical of black hole--black hole (BH--BH) or black hole--neutron star (BH--NS) binaries, up to redshift $z\approx 20$ and even higher. LIGO-Virgo-Kagra (LVK) has already detected 90 gravitational wave events to date. The most recent detections are presented in the GWTC-3 catalogue \citep{2021arXiv211103606T}.  Given that third-generation gravitational wave detectors will have high detection rates and redshift reach, and  will also be able to provide strong constraints on population properties thanks to the smaller uncertainties on the physical parameters of the GW events, \citep{2022A&A...667A...2S, 2022A&A...663A.156Y,2010arXiv1003.1386V} in this paper, we investigate the prospects of using the ET as a single instrument to reconstruct the functional form of the star formation rate (SFR) for a redshift range of $0 \leqslant z \leqslant 20$ for different compact binary populations.

The SFR density for Population (Pop) I stars for low redshifts is already very well constrained \citep{2014ARA&A..52..415M}, but there are large uncertainties for the first stars in the Universe, also known as Pop III stars. Pop III stars are expected to be the first sources of light and play a crucial role in the early cosmic evolution by producing the very first heavy elements \citep{2004ARA&A..42...79B, 2009Natur.459...49B}. Several observational methods have been used to probe this early stellar population \citep{2010MNRAS.402L..25K, toma2011population, meszaros2010population, 2011MNRAS.416.2760C, 2015MNRAS.449.3006M}, but direct observation of these Pop III stars remains to be achieved. The remnants of Pop III stars have been studied as possible sources of gravitational waves \citep{1984MNRAS.207..585B, 2004ApJ...608L..45B, 2006A&A...459.1001K, 2012A&A...541A.120K, 2014MNRAS.442.2963K, 2016MNRAS.460L..74H}.

There are some major differences between the evolution of Pop III stars and that of metal-polluted Pop I and II compact binaries (see \citet{2004ARA&A..42...79B} for detailed review). Pop III stars are expected to be more massive than Pop I stars \citep{2012ApJ...756...93H, 2012MNRAS.422..290S}, with masses of $10-100 M_{\odot}$, and so the Pop III star binaries are expected to evolve into BBHs. The main differences are as follows. (i) The initial conditions, such as initial mass function, initial binary mass ratio, and initial separations and eccentricities are different \citep{2017MNRAS.471.4702B}. (ii) The wind mass loss is very different. Heavy wind mass loss is seen for Pop I stars, smaller wind mass loss for Pop II stars, and almost no wind mass loss for Pop III stars. This in turn affects the binary separation evolution as wind mass loss widens the separation, meaning that Pop III stars are subject to more frequent binary interactions, such as Roche-lobe overflow (RLOF) and common envelope   (CE) interactions. The wind mass loss affects the neurton star and black hole mass, as there is more mass available for Pop III stars to form compact remnants (for a description of the effect of winds on mass, see \citet{2010ApJ...714.1217B}). (iii) There are also differences in radial evolution. Pop III stars have smaller radii than PopI and II stars \citep{2017MNRAS.471.4702B}. This leads to a smaller number of binary interactions for PopIII stars (RLOF and CE) as compared with PopI and II stars and thus counter balances the effect of wind loss. As there has not yet been direct detection of Pop III stars, the uncertainties due to different evolutionary parameters are still to be verified. It is therefore necessary to obtain strong constraints on the merger rate densities as a function of redshift in order to have a better understanding of the evolution of these first stars.

Assuming that the Pop III stars are formed by the collapse of dark matter halos, the SFR depends on multiple factors, including halo mass, reionisation history, metal enrichment in the intergalactic medium, and accretion rate. The ability of a primordial gas to cool and condense in the early Universe ---which in turn depends on the size of the `mini-halos'--- is one of the factors affecting star formation efficiency. The  star formation inside mini-halos, which are smaller in mass than the critical halo mass \citep{2003ApJ...592..645Y}, can be suppressed by ultraviolet background in the Lyman-Werner bands. Metal enrichment in the intergalactic medium ---where the main contribution is from supernova explosions--- determines when the formation of first stars will terminate. (\citet{2011A&A...533A..32D} explored many such factors in calculating the SFR of Pop III stars.) Therefore, any observational constraint on the SFR will be a crucial step in constraining formation scenarios.

For any given population of stars, three major questions regarding the SFR are as follows. (i) Firstly, we do not yet know when the star formation of Pop III stars terminated; (ii) secondly,  the redshift at which the SFR peaks in unknown; and finally, (iii) the functional form of SFR at high redshift remains to be determined. In this paper, we aim to find the answers to these questions with a given set of detections with single ET. We define the termination redshift as the redshift where the SFR drops to 1\% of its peak value. \citet{2019ApJ...886L...1V} specifically showed how detections of GWs from inspiraling BBHs by third-generation detectors can be used to measure the SFR of massive stars with high precision up to redshifts of approximately $10$ assuming that all sources come from galactic fields. The authors mention that they assume that the time-delay distribution is the same for all sources at all redshifts and neglected the dependence of this distribution on the mass and spin of the source. In order to estimate the redshift of the inspiralling binaries, these latter authors assume a three-detector 3G network. In this paper, we simulate multiple mock populations for compact binaries originating from Pop I+II (field binaries) and Pop III stars using a more realistic distribution of time delay to construct the mock population. While estimating the SFR we make no assumptions about the originating population of the compact binary or about the functional form of the SFR. We estimate the parameters of the compact binaries and the SFR with ET as a single instrument.

\section{Plan of the paper}

In \citet{2021PhRvD.104d3014S}(SB1 hereafter), we developed an algorithm to break the chirp mass--redshift degeneracy in the detected GW signal from the coalescence of a compact binary with ET as a single instrument, and thus to estimate the parameters of the merging compact binary. We estimated the area of localisation, chirp mass, redshift, and mass ratios by estimating their posterior distribution for short-duration GW signals from inspiraling compact binary systems. In the subsequent work in the series (\cite{2022PhRvD.106l3014S}; SB2 hereafter), we further developed the algorithm, taking into account the effect of the rotation of the Earth on the antenna pattern function in order to analyse the long-duration signals from coalescing low-mass compact binary systems. We used this algorithm to further analyse realistic populations of compact binary systems originating from Pop I+II, Pop III, and globular cluster (GC) populations in (\cite{2022A&A...667A...2S}; S22 hereafter). In S22, we concluded that ET as a single instrument is capable of detecting and distinguishing different compact binary populations separated in chirp mass--redshift space. We also estimated the merger rate density and found that, although our estimates for Pop I+II and GC populations are in good agreement with the true merger rate density of the respective populations, the deviation from the true value is much larger in the case of Pop III, especially for higher redshifts of $z>7$ (see Fig. 7 in S22). In the present work, we use an improved version of the algorithm
to estimate the parameters of the merging compact binaries.

In \S \ref{pop_description} we describe the population models and the SFRs used to generate our mock populations. We discuss estimations of the parameters of the detected compact binaries in \S \ref{PE_ET}. To obtain an estimate of the SFR of a given population, it is crucial to have a correct assessment of the merger rate density as a function of redshift. In order to estimate the merger rate density accurately, we calculate the detection efficiency. Taking into account the fraction of binaries that do not cross the detection threshold, we reconstruct the true merger rate density of the population. This is described in detail in \S \ref{plan_pop_mer}. Subsequently, assuming a functional form for the delay time, which is the time from the formation of the stars to the merger of the compact binaries formed from those stars, we proceed to reconstruct the SFR of the population. This process is described in detail in \S \ref{sfr_rate}. We discuss our conclusions in \S \ref{conc}.

\section{Mock source catalogue}\label{pop_description}

As mentioned in the previous section, in S22 we showed that ET as a single instrument is capable of detecting and distinguishing different compact binary populations separated in chirp mass--redshift space. Based on this earlier result, in the present work, we simulate three mock populations for compact binaries originating from Pop I+II and Pop III stars: Two mock populations are generated that consist of only compact binaries of Pop III stars using two realistic star formation densities evolving over redshift. We assume that compact binaries originating from these populations are clearly distinguishable from other populations. One mock population is generated consisting of compact binaries from both Pop I+II and Pop III stars, allowing us to estimate the SFR without assuming the ability to  distinguish between these populations.

We use the FS1 model of the population of compact binary systems from the first, metal-free  Pop III stars generated by \citet{2017MNRAS.471.4702B}. The initial conditions for generating this model are based on the models obtained by \cite{2016MNRAS.456..223R} using N-body simulation, and assuming the formation of Pop III stars from a mini halo of $\sim 2000$ AU in  size \citep{2013MNRAS.433.1094S}. The number density of the gas medium is chosen to be $10^6\rm{cm}^{-3}$, following the parameters specified by \cite{2013MNRAS.433.1094S}. The details of this population of metal-free binaries, such as initial mass function, mass ratio, orbital separations, and eccentricities, are described in \cite{2017MNRAS.471.4702B}.  

\citet{2011A&A...533A..32D} considered two populations of Pop III stars: (i) Pop III.1 stars, which are the first-generation stars formed from initial conditions determined cosmologically, and (ii) Pop III.2 stars, which are zero-metallicity stars that formed from a primordial gas, influenced by an earlier generation of stars. Pop III.2 stars are expected to form in an initially ionised gas \citep{2006MNRAS.366..247J, 2007ApJ...663..687Y} and are thought to be less massive $(\sim 40 - 60 M_{\odot})$ than Pop III.1 stars $(\sim 1000M_{\odot})$. \citet{2011A&A...533A..32D} calculated the SFR for Pop III.1 and Pop III.2 using three different values of the galactic wind ---namely $v_{\rm wind} = 50, 75, 100$km/s--- in order to incorporate the effect of metal enrichment by galactic winds, and two different star formation efficiency values, $f_{\star} = 0.001$ and $f_{\star} = 0.1$. In the present paper, we consider three SFRs calculated by \citet{2011A&A...533A..32D} for Pop III.2 stars: (i) $v_{\rm wind} = 50$ km/s and $f_{\star} = 0.001,$ (ii) $v_{\rm wind} = 100$ km/s and $f_{\star} = 0.001$, and (iii) a very optimistic case with $v_{\rm wind} = 50$ km/s and $f_{\star} = 0.1$. We refer to these SFRs as SFR1, SFR2, and SFR3, respectively. 

\begin{figure}
\centering
\includegraphics[width=\columnwidth]{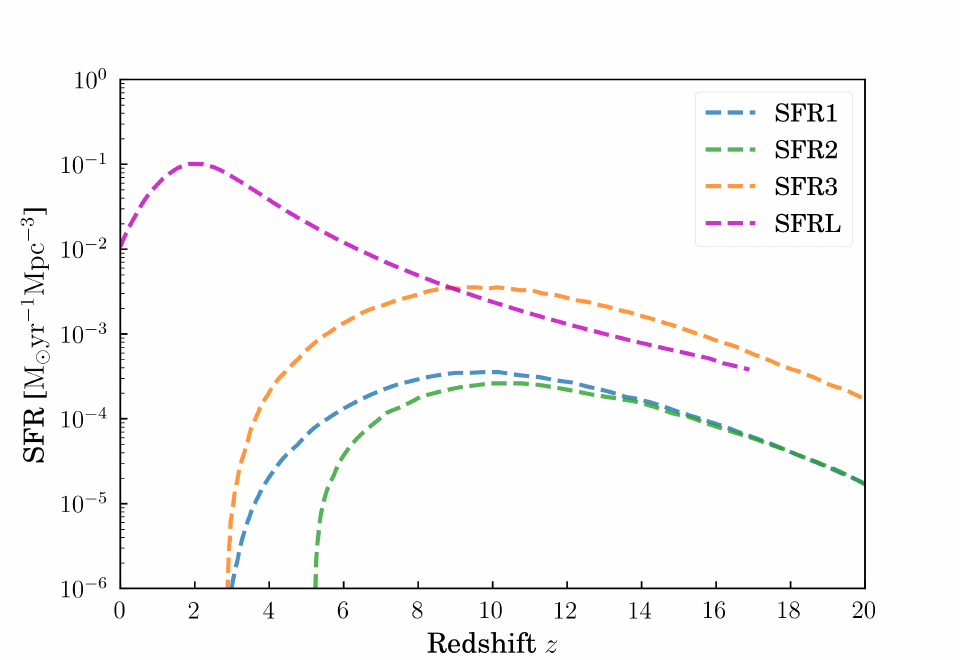}
\caption{Star formation rates used to construct the mock populations in this analysis. SFRs for Pop III are adopted from \citet{2011A&A...533A..32D} and the SFR for Pop I+II is taken from \citet{2020A&A...636A.104B}.}
\label{fig:sfr_all}
\end{figure}

These SFRs are shown in Figure \ref{fig:sfr_all} and we see that the major effect of metal enrichment by galactic winds in the case of SFR1 and SFR2 is manifested in the termination (the redshift where the SFR drops to 1\% of its peak value) of these SFRs. SFR2 has greater galactic wind velocity as compared to SFR1 and we see that SFR2 terminates at $ z \sim 5.3,$ whereas SFR1 terminates at a much later time, at $z \sim 3.2$. Therefore, any constraint on the termination redshift will be helpful in providing information about the formation scenarios of these first stars (for details see \cite{2004ARA&A..42...79B}). While the metal enrichment via galactic winds is the same for SFR1 and SFR3, the increased star formation efficiency, $f_{\star}$, leads to an overall increase in the absolute value of SFR as a function of redshift.

For the Pop I+II compact binaries, we use the M30B generated by \citet{2020A&A...636A.104B} and an upgraded version of the population synthesis code StarTrack  \citep{2002ApJ...572..407B, 2008ApJS..174..223B}. These authors generated multiple binary stellar evolution models consistent with LVK O1/O2 merger rates for BBH and BNS mergers \citep{PhysRevX.9.031040}. The input physics for the M30 model is summarised in Table 2 of \citet{2020A&A...636A.104B}. The extension `B'  specifies the models that do not allow a common envelope with Hertzsprung gap donors. The M30B model used in this analysis was generated using the SFR specified in Eq. (16) in \citet{2020A&A...636A.104B}, taking into account the evolution of metallicity with redshift using Eq. (18) in \citet{2020A&A...636A.104B}. We refer to this SFR as the SFRL and show it in Figure \ref{fig:sfr_all}. We construct the following three mock populations for our analysis:

\textbf{Mock 1}: Population consisting of only Pop III compact binaries, constructed using the model FS1 and SFR1.

\textbf{Mock 2}:  Population consisting of only Pop III compact binaries, constructed using the model FS1 and SFR2.

\textbf{Mock 3}: Population consisting of both Pop I+II and Pop III compact binaries, constructed with the model M30B for Pop I+II using the SFRL, and with the model FS1 for Pop III using SFR3. We used SFR3 here to have a non-negligible merger rate of Pop III binaries as compared to Pop I+II compact binaries.
 
The expected merger rates per year for Mock 1 and Mock 2 are 2429 and 1450, respectively, and as we generated 80\,725 for Mock 1 and 68\,500 compact binaries for Mock 2, these correspond to an observation time of $\sim 33$ yr and $\sim 47$ yr, respectively. In the case of Mock 3, we generated 162\,106 compact binaries, which corresponds to $\sim 0.5$ yr of observation for Mock 3.

\subsection{Construction of the mock populations}\label{mock_pop}

For both M30B and FS1 compact binaries, we use the population available on the StarTrack\footnote{http://www.syntheticuniverse.org/} website. For the FS1 model, we only consider the following data for each binary: (i) the masses of the merging compact objects $m^i_{1,2}$, and (ii) the delay time between the formation of the binary at zero-age main sequence (ZAMS)  and its coalescence, $t_{\rm del}^i$. For compact binaries of the M30B model, 
we use the following data for each binary: (i) the masses of the merging compact objects $m^i_{1,2}$, (ii) the merger redshift, and (ii) the merger rate density for each binary in the observer frame of reference.

For a ZAMS binary with masses $m_{1,2}$, delay times $t_{del}$, and metallicity $Z$ formed at cosmic time $t_{\rm ini} = t_{\rm obs} - t_{\rm del}$, corresponding to redshift $z_{\rm ini} = z(t_{\rm ini})$, the delay time $t_{\rm del} = t_{\rm evol} + t_{\rm merg}$. The $t_{\rm evol}$ is the time of evolution from ZAMS to the formation of a compact binary system; $t_{\rm merg}$ is the time till the merger; and $t_{\rm obs}$ is the cosmic time at which the compact binary is observed to merge. Then, for a binary system `$b$', the merger rate density per unit redshift as a function of redshift is given as:
\begin{equation}\label{Rofi}
\mathcal{R}_b(z) = \frac{1}{(1+z)}\frac{dV}{dz}\left(\frac{{\rm{SFR}}(z_{i, ini}),Z_b)}{M_{\rm sim}}\right),
\end{equation}
where $M_{\rm sim}$ is the total mass of all stars that must accompany the stellar evolution, leading to formation of compact object binaries. These include the binaries and the single stars. $M_{\rm sim}$ for M30B and FS1 is $2.8\times 10^8 M_{\odot}$ \cite{2020A&A...636A.104B} and $3.5 \times 10^9 M_{\odot}$ \citep{2017MNRAS.471.4702B}, respectively. 

Equation (\ref{Rofi}) gives the redshift dependence of the merger rate density of each binary, and so the probability density of a merger of a type $b$ to happen in the  Universe at redshift $z$ is proportional to $\mathcal{R}_b(z)$:
\begin{equation}\label{probzi}
P(b,z) \propto \mathcal{R}_b(z),
\end{equation}
which is  discrete in the index $b$ and continuous in $z$. The probability distribution can be obtained by normalisation. We construct the mock populations using different SFRs, as mentioned above. To generate a mock population, we sample random binaries from the distribution given by Equation (\ref{probzi}) and to each compact binary system drawn in this manner we assign random values to the four angles: the right ascension $\alpha$, the angle of declination $\delta$, the polarisation angle $\psi,$ and the inclination angle $\iota$ of the binary with respect to the direction of observation. The values of $\cos\delta, \alpha/ \pi$, $\cos \iota,$ and $\psi/ \pi$ are chosen to be uncorrelated and distributed uniformly over the range $[-1,1]$. In the following section, we describe the method to estimate the parameters of merging compact binaries detected with ET as a single instrument.

\begin{figure*}[!]
\centering
\subfloat[\label{fig:50km_chm_med}]{\includegraphics[width=\columnwidth]{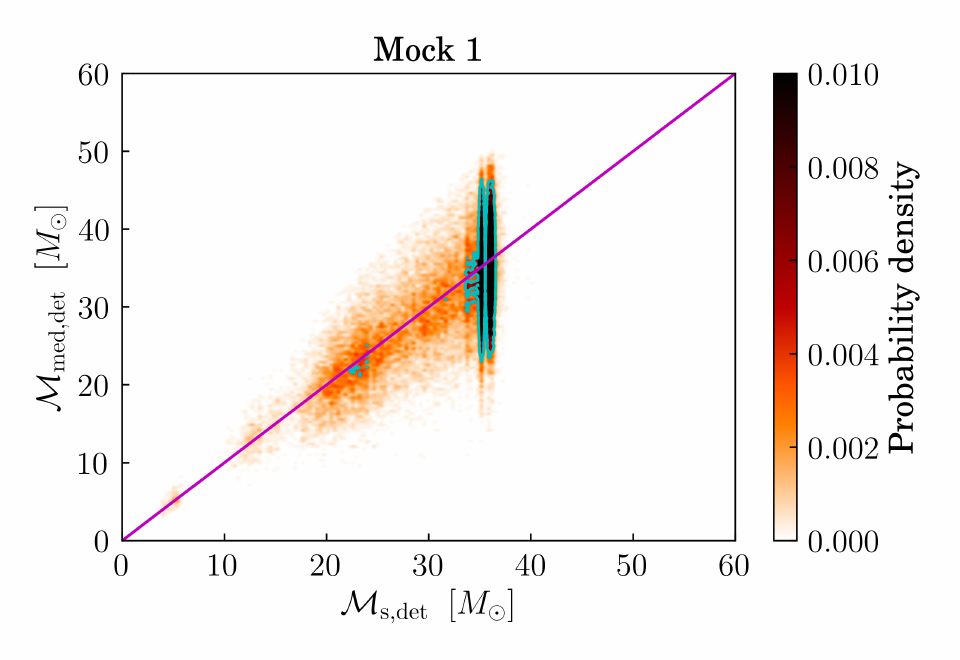}}\subfloat[\label{fig:50km_red_med}]{\includegraphics[width=\columnwidth]{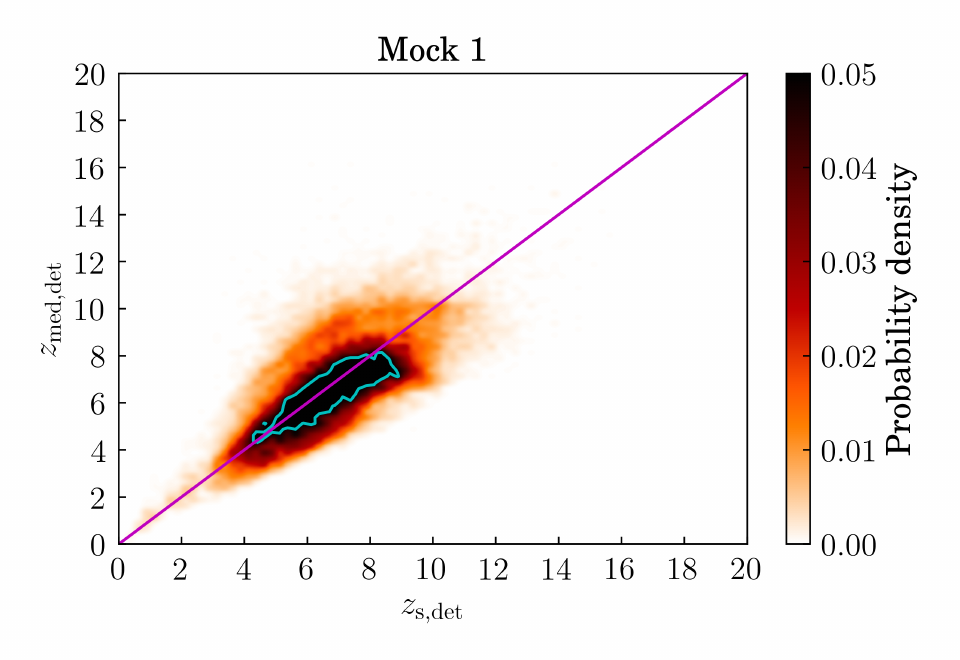}}\\
\subfloat[\label{fig:100km_chm_med}]{\includegraphics[width=\columnwidth]{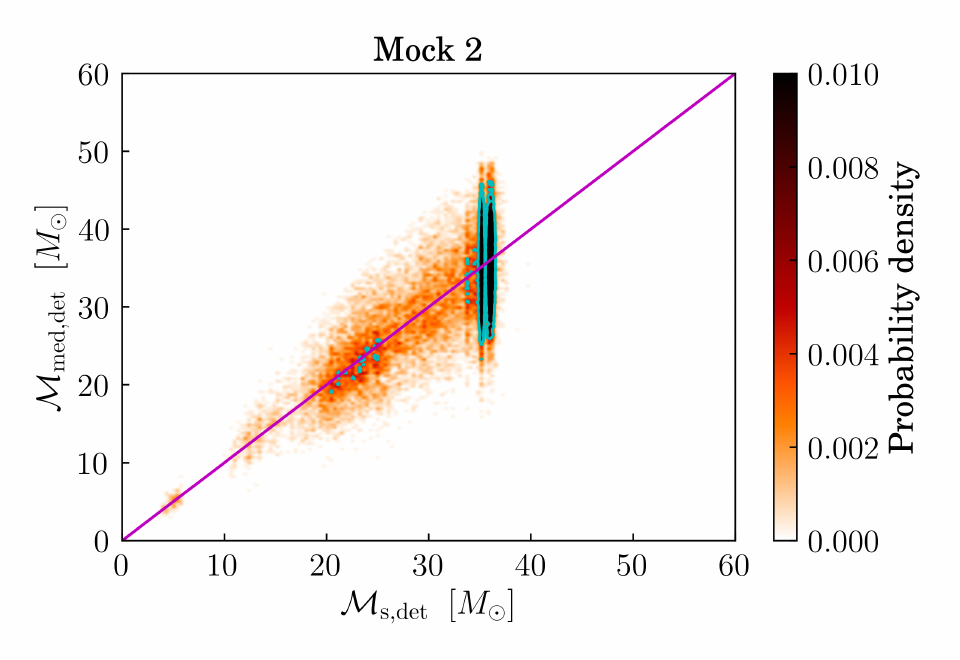}}\subfloat[\label{fig:100km_red_med}]{\includegraphics[width=\columnwidth]{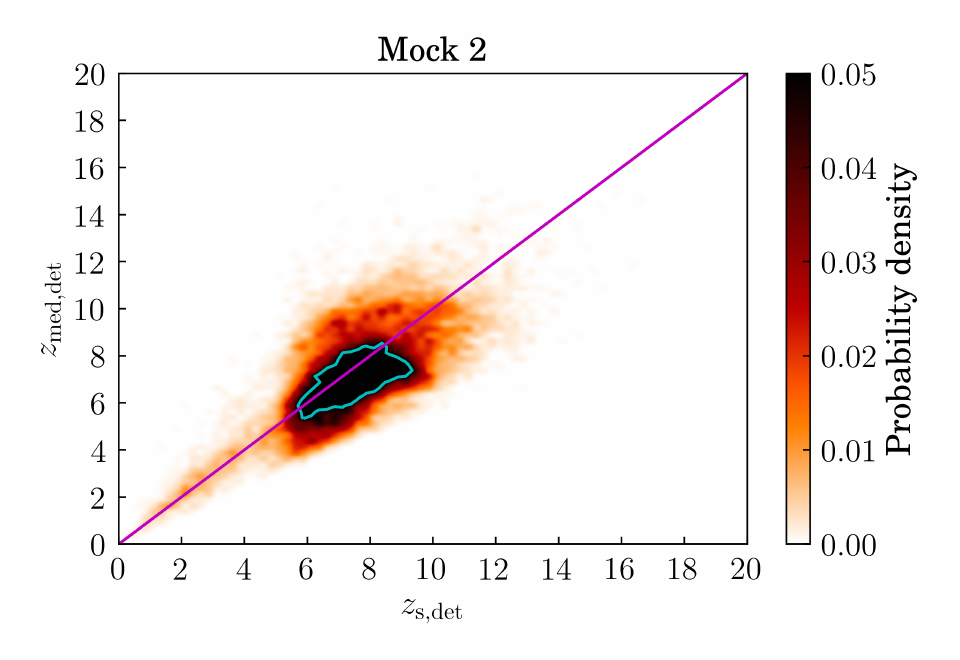}}\\
\subfloat[\label{fig:mixed_chm_med}]{\includegraphics[width=\columnwidth]{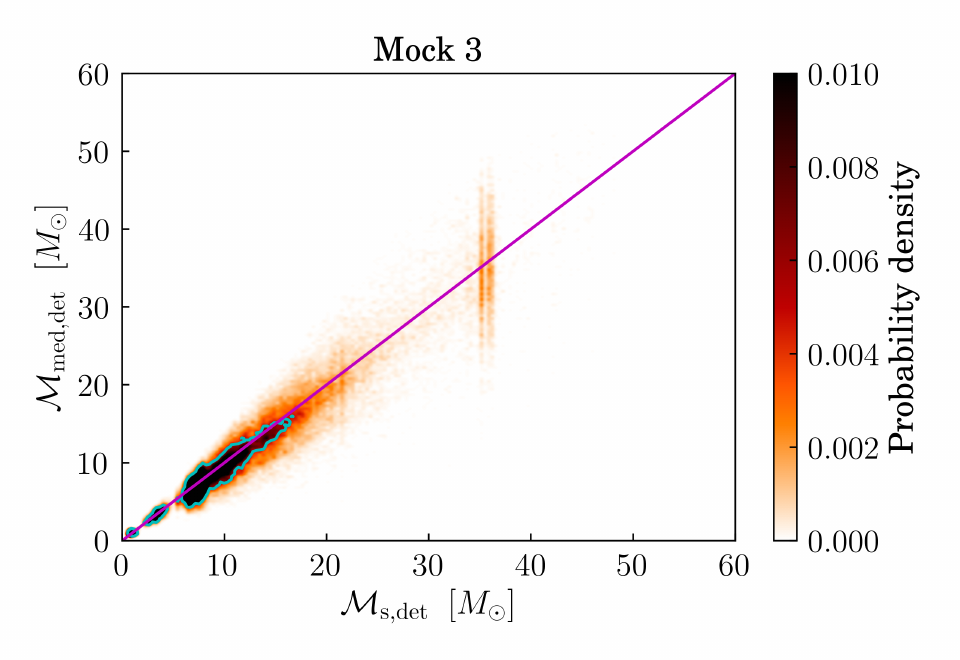}}\subfloat[\label{fig:mixed_red_med}]{\includegraphics[width=\columnwidth]{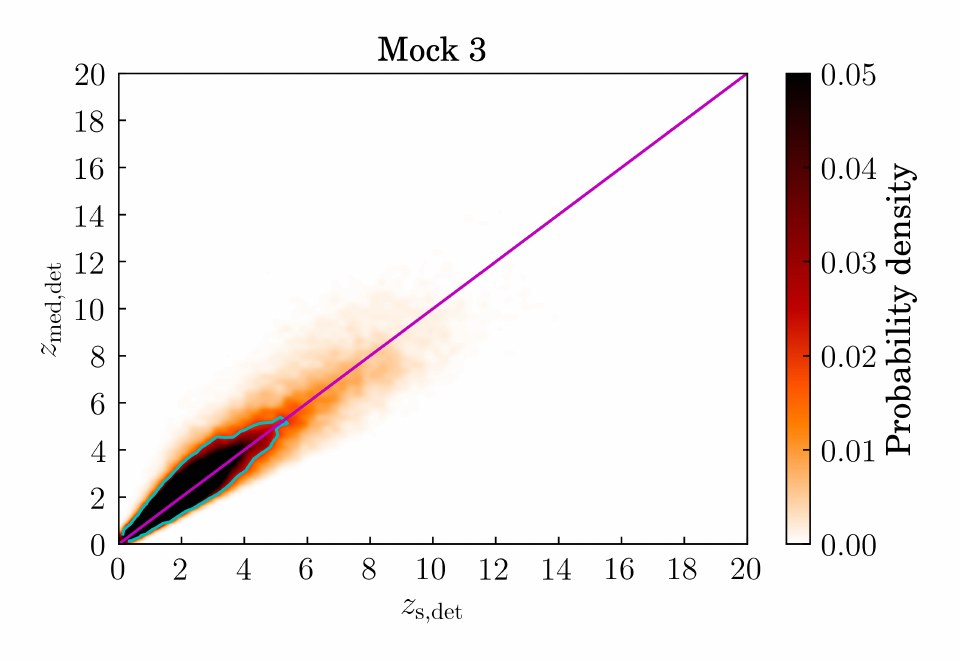}}\\
\caption{Estimation of parameters. Density distribution of the median of the estimated posterior with respect to the true values of the parameters: Chirp mass $\mathcal{M}$ (left) and redshift $z$ (right) of the detected compact binary sources. Top panel: Mock 1. Middle panel: Mock 2. Bottom panel: Mock 3. The magenta line is a reference for equal values of true and estimated parameters. The blue contour encloses the 90\% probability region.}
\label{fig:injvsmedian}
\end{figure*}

\section{Estimation of the parameters of compact binaries with ET }\label{PE_ET}

For a merging compact binary of chirp mass $\mathcal{M}$, located at a luminosity distance $D_{\rm L}$, the two polarisations of GWs  at time $t < t_c$  according to the theory of general relativity  \citep{Findchirp} are

\begin{subequations}\label{hpluscross_ant}

\begin{equation}
\begin{split}
    h_{+}(t) = -\frac{1+\cos^{2}\iota}{2}\left(\frac{G \mathcal{M} }{c^2D_{\rm L}}\right)\left(\frac{t_c-t}{5G\mathcal{M} /c^3}\right)^{-1/4}\\
    \times \cos\left[2\Phi_c + 2\Phi \left(t-t_c ; M,\mu\right)\right],
\end{split}
\end{equation}

\begin{equation}
\begin{split}
    h_{\times}(t) = -\cos\iota \left(\frac{G \mathcal{M}}{c^2D_{\rm L}}\right)\left(\frac{t_c-t}{5G\mathcal{M}/c^3}\right)^{-1/4}\\
    \times \sin\left[2\Phi_c + 2\Phi \left(t-t_c ; M,\mu\right)\right],
\end{split}
\end{equation}
\end{subequations}
where $c$ is the speed of light, $G$ is the gravitational constant, $\mu$ is the reduced mass of the binary system, $\iota$ is the angle of inclination of the orbital plane of the binary system with respect to the observer, and $\Phi \left(t-t_c ; M,\mu\right)$ is the orbital phase of the binary system. For a binary system with component masses $m_1$ and $m_2$, the chirp mass is $\mathcal{M} = (m_1m_2)^{3/5}/M^{1/5}$, and the total mass is $M = m_1+m_2$. The quantities $t_{c}$ and $\Phi_{c}$ are the time and phase, respectively, at the termination of the waveform \citep{Findchirp}.  The strain  $h(t)$ generated in a detector due to this waveform is

\begin{equation}\label{h_t_ant}
    h(t)= F_+ h_+(t+t_c-t_0) + F_{\times} h_\times(t+t_c-t_0),
\end{equation}
where $t_0$ is the time of coalescence in the detector frame, and so $(t_0 - t_c)$ is the travel time from the source to the detector. $F_+, F_\times$ are the antenna response functions of one of the three detectors in ET. The strain can be rewritten as 

\begin{equation}\label{ht}
\begin{split}
    h(t)= -\left(\frac{G\mathcal{M}}{c^2}\right)\left(\frac{\Theta}{4D_L}\right)\left(\frac{t_0-t}{5G\mathcal{M}/c^3}\right)^{-1/4}\\
    \times \cos\left[2\Phi_0 + 2\Phi \left(t-t_c ; M,\mu\right)\right]
\end{split}
\end{equation}
by substituting the values of the two polarisations from Equation (\ref{hpluscross_ant}) into Equation (\ref{h_t_ant}):
The functions $\Theta$ and the phase $\Phi_{c}$ in Equation (\ref{ht}) are functions of the antenna response functions $F_+$ and $F_\times$ and the angle of inclination $\iota,$ and are defined as
\begin{equation}\label{theta}
    \Theta\equiv 2 \left[F_{+}^{2}\left(1+\cos^{2}\iota\right)^{2} + 4F_{\times}^{2}\cos^{2}\iota\right]^{1/2},
\end{equation}
such that  $0<\Theta<4$ and
\begin{equation}\label{snr_phase}
    2\Phi_0 = 2\Phi_c - \arctan\left(\frac{2F_\times\cos\iota}{F_+\left(1 + \cos^2\iota\right)}\right).
\end{equation}
Assuming that the three ET detectors have identical noise, the signal to noise ratio (S/N), $\rho_j$ , for $j = (1,2,3)$ in each of the three ET detectors is given as \citep{TaylorGair2012}
\begin{equation}\label{snr}
\rho_j \approx 8 \Theta_j \frac{r_{0}}{D_{\rm L}}
\left(\frac{\mathcal{M}_{z}}{\mathcal{M}_{\rm BNS}}\right)^{5/6}\sqrt{\zeta\left(f_{\rm max}\right),} 
\end{equation}
where  the redshifted chirp mass $\mathcal{M}_{z}= (1+z)\mathcal{M} $ and the reference mass $\mathcal{M}_{\rm BNS}\approx 1.218 M_\odot $  is the chirp mass of an equal mass binary with components of $1.4 M_{\odot}$ each. The function $\zeta\left(f_{\rm max}\right)$ is defined as

\begin{equation}\label{zetafunc}
\zeta\left(f_{\rm max}\right) = \frac{1}{x_{7/3}}\int^{2f_{max}}_{1}\frac{df \left( \pi M_{\odot}\right)^{2}}{\left(\pi f M_{\odot}\right)^{7/3}S_{h}\left(f\right)},
\end{equation}
where $S_{h}\left(f\right)$ is the power spectral density (PSD) with the ET-D noise curve \citep{2011CQGra..28i4013H} for the ET-D configuration and,

\begin{equation}\label{x_7_3}
x_{7/3} = \int^{\infty}_{1}\frac{df \left( \pi M_{\odot}\right)^{2}}{\left(\pi f M_{\odot}\right)^{7/3}S_{h}\left(f\right)},
\end{equation}

\begin{equation}\label{detreach}
r^{2}_{0} = \frac{5}{192 \pi}\left(\frac{3 G}{20}\right)^{5/3}x_{7/3}\frac{M^{2}_{\odot}}{c^{3}},
\end{equation}

\begin{equation}\label{fmax}
f_{\rm max} = 785\left(\frac{M_{\rm BNS}}{M(1+z)}\right)  \;\rm Hz,
\end{equation}
where $r_0$ is the characteristic distance sensitivity and $f_{\rm max}$ is the frequency at the end of the inspiral phase. For the ET triangular configuration consisting of three detectors, the combined effective S/N is given as
\begin{equation}\label{snreff}
\rho_{\rm eff} = 8 \Theta_{\rm eff} \frac{r_{0}}{D_{\rm L}}\left(\frac{\mathcal{M}_{z}}{1.2 M_{\odot}}\right)^{5/6}\sqrt{\zeta\left(f_{\rm max}\right)}, 
\end{equation}
where the effective antenna response function  $\Theta_{\rm eff}$ is

\begin{equation}\label{thetaeff}
\Theta_{\rm eff} = \left(\Theta_{1}^{2} + \Theta_{2}^{2} + \Theta_{3}^{2}\right)^{1/2}.
\end{equation}

\subsection{Summary of the algorithm used in SB2}
In SB2, we estimate the parameters of the merging compact binaries, taking into account the change in the antenna pattern with the rotation of the Earth in order to analyse the long-duration GW signals. We assume the location of the ET detector to be the Virgo site, and  we analyse the signal every 5 minutes. The observed GW frequency  $f^{\rm obs}_{\rm gw}$ is calculated using Equation (4.195) in \citet{Maggiore:1900zz}:

\begin{equation}\label{f_gw}
f^{\rm obs}_{\rm gw} = \frac{1}{\pi} \left( \frac{5}{256}\frac{1}{\tau_{\rm obs}}  \right)^{3/8} \left( \frac{G\mathcal{M}_{z}}{c^3}  \right)^{-5/8},
\end{equation}
where $\tau_{\rm obs}$ is the time to coalescence, which is measured in the observer's frame. The minimum frequency given the detection sensitivity of the detector and the frequency $f_{\rm max}$ determines the limit on $\tau_{\rm obs}$ in the detection band. For $\tau_{i-1}$ and $\tau_{i}$, which are the initial and the final values of $\tau_{\rm obs}$, respectively, for the $i^{th}$ segment, the corresponding values $f_{i-1}, f_{i}$ of $f^{\rm obs}_{\rm gw}$ will be
\begin{equation}
f_{i-1} = \frac{1}{\pi} \left( \frac{5}{256}\frac{1}{\tau_{i-1}}  \right)^{3/8} \left( \frac{G\mathcal{M}_{z}}{c^3}  \right)^{-5/8}
,\end{equation}
and
\begin{equation}
f_{i} = \frac{1}{\pi} \left( \frac{5}{256}\frac{1}{\tau_{i}}  \right)^{3/8} \left( \frac{G\mathcal{M}_{z}}{c^3}  \right)^{-5/8}.
\end{equation}
The S/N for the $i^{th}$ segment in the $j^{th}$ detector can be written using equation (\ref{snr}), as

\begin{equation}\label{snr_seg}
\rho^i_j \approx 8 \Theta^i_j \frac{r_{0}}{D_{L}}
\left(\frac{\mathcal{M}_{z}}{\mathcal{M}_{\rm BNS}}\right)^{5/6}\sqrt{\zeta^i\left(f_{i-1}, f_{i}\right), }
\end{equation}
where

\begin{equation}\label{zetafunc_seg}
\zeta^i\left(f_{i-1}, f_{i}\right) = \frac{1}{x_{7/3}}\int^{f_i}_{f_{i-1}}\frac{df \left( \pi M_{\odot}\right)^{2}}{\left(\pi f M_{\odot}\right)^{7/3}S_{h}\left(f\right)},
\end{equation}
and 

\begin{equation}\label{theta_seg}
    \Theta^i_j\equiv 2 \left[(F^i_{+})^{2}\left(1+\cos^{2}\iota\right)^{2} + 4(F^i_{\times})^{2}\cos^{2}\iota\right]^{1/2}_j  ,
\end{equation}
where $F^i_{+}$ and $F^i_{\times}$ are the antenna response functions for the $j^{th}$ detector in the $i^{th}$ segment. The effective S/N for the $i^{th}$ segment is

\begin{equation}\label{snreff_seg}
\rho^i_{\rm eff} = 8 \Theta^i_{\rm eff} \frac{r_{0}}{D_{\rm L}}\left(\frac{\mathcal{M}_{z}}{1.2 M_{\odot}}\right)^{5/6}\sqrt{\zeta^i\left(f_{i-1}, f_{i}\right),} 
\end{equation}
where

\begin{equation}
(\rho^i_{\rm eff})^2 = (\rho^i_1)^2 + (\rho^i_2)^2 + (\rho^i_3),^2 
\end{equation}
and the function  $\Theta^i_{eff}$ is
\begin{equation}\label{thetaeff_seg}
(\Theta^i_{\rm eff})^2 = (\Theta^i_{1})^{2} + (\Theta^i_{2})^{2} + (\Theta^i_{3})^{2}.
\end{equation}
We assume that the observables, in the case of the detection of a coalescing binary system, are:
(a) the three S/Ns $\rho^i_j$ for the $i^{th}$ segment of the signal, (b) the phase $\Phi^i_{o,j}$ for $j = (1,2,3)$ for each of the three ET detectors in the $i^{th}$ segment of the signal, (c) the GW frequency at the start and end of each segment of the detected signal, (d) the redshifted chirp mass $\mathcal{M}_{z}$, and (e) the frequency at the end of the inspiral, corresponding to the innermost stable circular orbit, $f_{\rm max}$. We assume the measurement errors on the S/Ns to be Gaussian, such that the standard deviations for $\rho^i_j$ and $\Phi^i_{j}$ are $\sigma_{\rho}=1$ and $\sigma_{\Phi}=\pi/\rho$, respectively. This is a conservative assumption as compared to the errors on the S/Ns of the GW detections mentioned in GWTC-2 and GWTC-3 \citep{2021PhRvX..11b1053A,2021arXiv211103606T}.

One of the main building blocks of the algorithm in SB2 is that we use the ratios of S/N in each segment in order to constrain $\Theta^i_{\rm eff}$ (see Sec IV of SB2):
\begin{equation}\label{theta-ratios}
     \rho^i_{21} = \frac{\Theta^i_2}{\Theta^i_1}\equiv\Theta^i_{21} \ \ \ {\rm and}\ \ \ \ \   \rho^i_{31} = \frac{\Theta^i_3}{\Theta^i_1}\equiv \Theta^i_{31},
\end{equation}
where $\rho^i_{21} \equiv \rho^i_2/\rho^i_1$ and $\rho^i_{31} \equiv \rho^i_3/\rho^i_1$ in the $i^{th}$ segment. We then proceed to constrain the source-dependent quantity $\Lambda$ (defined in Equation 37 of SB2) for each segment in order to constrain the binary parameters such as chirp mass, total mass, mass ratio, redshift, and luminosity distance:

\begin{equation}\label{lambda}
    \Lambda \equiv \left(\frac{8r_{0}}{D_{\rm L}}\left(\frac{\mathcal{M}_{z}}{\mathcal{M}_{\rm BNS}}\right)^{5/6} \right)^{-1}  \approx \frac{\Theta^i_{\rm eff} \sqrt{\zeta^i\left(f_{i-1}, f_{i}\right)}}{\rho^i_{\rm eff}}.
\end{equation}
With this algorithm, we find that the chirp masses are overestimated, while the redshift is underestimated. This was also seen in the estimates of the binary parameters for Pop III binaries carried out in S22. The origin of this `bias' is the probability distribution of $\Theta$ (see Figure 13 of SB2) given the antenna pattern functions of the triangular configuration of ET (see Appendix of SB2 for detail).

\subsection{Modification in the algorithm used in SB2}\label{algo_modif}
By comparing both sides of Equation (\ref{lambda}), we find a function $\mathcal{F}(\rho^i_{21}, \rho^i_{31}),$ such that
\begin{equation}\label{fit_def}
        \mathcal{F}(\rho^i_{21}, \rho^i_{31}) = \frac{\Lambda_{\rm s}}{\Lambda_{\rm med}} ,
\end{equation}
where the subscript `s' denotes the true value of $\Lambda$ from the actual source parameters,
\begin{equation}\label{lambda_s}
        \Lambda_{\rm s} = \left(\frac{8r_{0}}{D_{\rm L}}\left(\frac{\mathcal{M}_{z}}{\mathcal{M}_{\rm BNS}}\right)^{5/6} \right)^{-1},
\end{equation}
and $\Lambda_{\rm med}$ is the median of the estimated posterior distribution of $\Lambda$ in the $i^{th}$ segment of the signal:
\begin{equation}\label{lambda_med}
        \Lambda_{\rm med} = \left( \frac{\Theta^i_{\rm eff} \sqrt{\zeta^i\left(f_{i-1}, f_{i}\right)}}{\rho^i_{\rm eff}}\right)_{\rm median}.
\end{equation}
We approximate $\mathcal{F}(\rho^i_{21}, \rho^i_{31})$ to a fit of the form
\begin{equation}\label{fit_formula}
    \mathcal{F}(\rho^i_{21}, \rho^i_{31}) \approx 1+ \frac{\mathcal{R}(\rho^i_{21}, \rho^i_{31},1,1,0.022)}{1.36\times 10^6} + \frac{\mathcal{R}(\rho^i_{21}, \rho^i_{31},1,1,0.7),}{8}
\end{equation}
where $\mathcal{R}$ is the Marr or Mexican hat function, and is given as
\begin{equation}
\begin{split}
        \mathcal{R}(x,y, x_0, y_0, \sigma) = \frac{1}{\pi\sigma^4} & \left(1-\frac{1}{2} \left(\frac{(x-x_0)^2+(y-y_0)^2}{\sigma^2}\right)\right). \\
        & \times e^{-\frac{(x-x_0)^2+(y-y_0)^2}{2\sigma^2}}
\end{split}
\end{equation}
A more detailed description of this fit is described in \S \ref{sec:bias_construct}. We include this function $\mathcal{F}(\rho^i_{21}, \rho^i_{31})$ in Equation 39 in SB2 as prior information, and continue the rest of the analysis to estimate the parameters of a given binary as described in SB2. In this work, we fix the detection  at a threshold value of accumulated effective S/N $\rho_{\rm eff}>8,$ and the S/N for $i^{th}$ segment in the $j^{th}$ detector $\rho^i_j> 3$ in at least one segment for $j = (1,2,3),$ corresponding to the three ET detectors comprising the single ET. We then use this improved algorithm to estimate the chirp mass, total mass, and redshift of a detected compact binary system from the three mock populations. 

We consider three populations of compact binaries originating from Pop III stars and merging within a Hubble time based on three different SFRs. The construction of all three populations is described in \S \ref{mock_pop}. For a given population of compact binaries, the true values of the chirp mass, total mass, and redshift of these `sources' are represented as $\mathcal{M}_{\rm s,mock}$, $M_{\rm s, mock}$, and $z_{\rm s, mock}$, respectively. A binary source is considered as `detected' if it crosses a detection threshold set on the S/N as mentioned in \S \ref{algo_modif}. The chirp mass, total mass, and redshift of these detected sources are denoted $\mathcal{M}_{\rm s,det}$, $M_{\rm s,det}$, and $z_{\rm s, det}$, respectively. The posterior probability distribution for the chirp mass and redshift for each of these detected sources is estimated using the algorithm described above (see Fig. 5 in SB2 for an example). The median values of these estimated posterior probability distributions of chirp mass, total mass, and redshift for each detected compact binary source are represented as $\mathcal{M}_{\rm med,det}$, $M_{\rm med,det}$, and $z_{\rm med, det}$, respectively.

\begin{figure*}[!]
\centering
\subfloat[\label{fig:50km_rate}]{\includegraphics[width=0.33\textwidth]{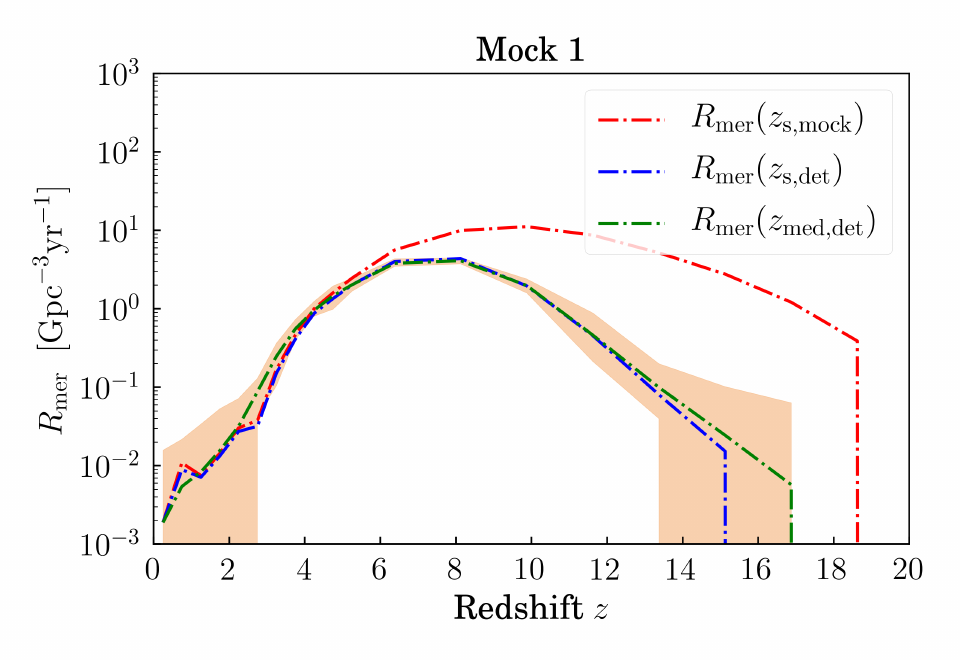}}
\subfloat[\label{fig:50km_rel}]{\includegraphics[width=0.33\textwidth]{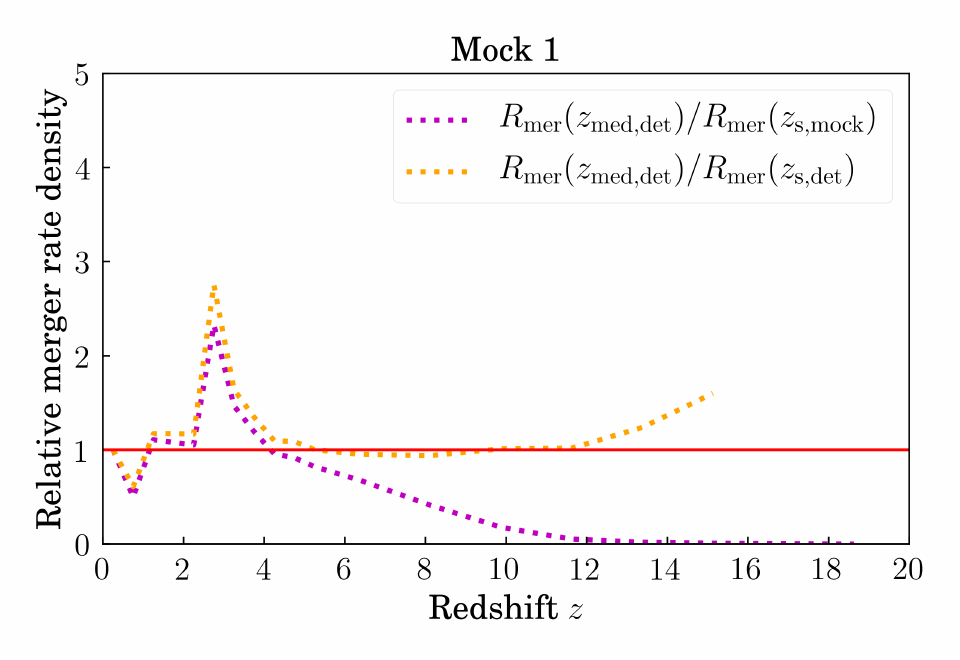}}\subfloat[\label{fig:50km_cummu}]{\includegraphics[width=0.33\textwidth]{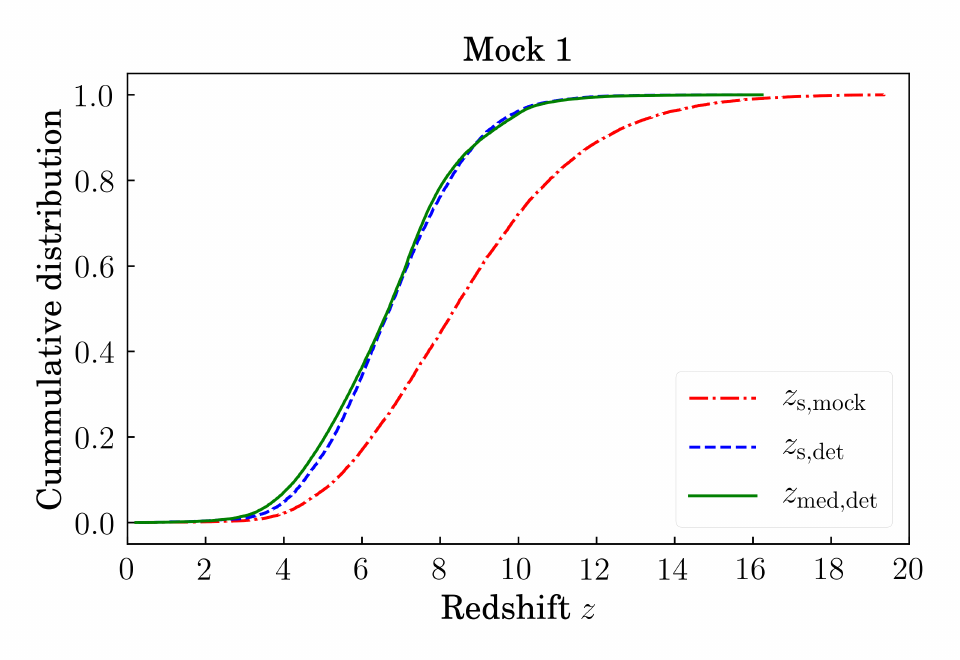}}\\
\subfloat[\label{fig:100km_rate}]{\includegraphics[width=0.33\textwidth]{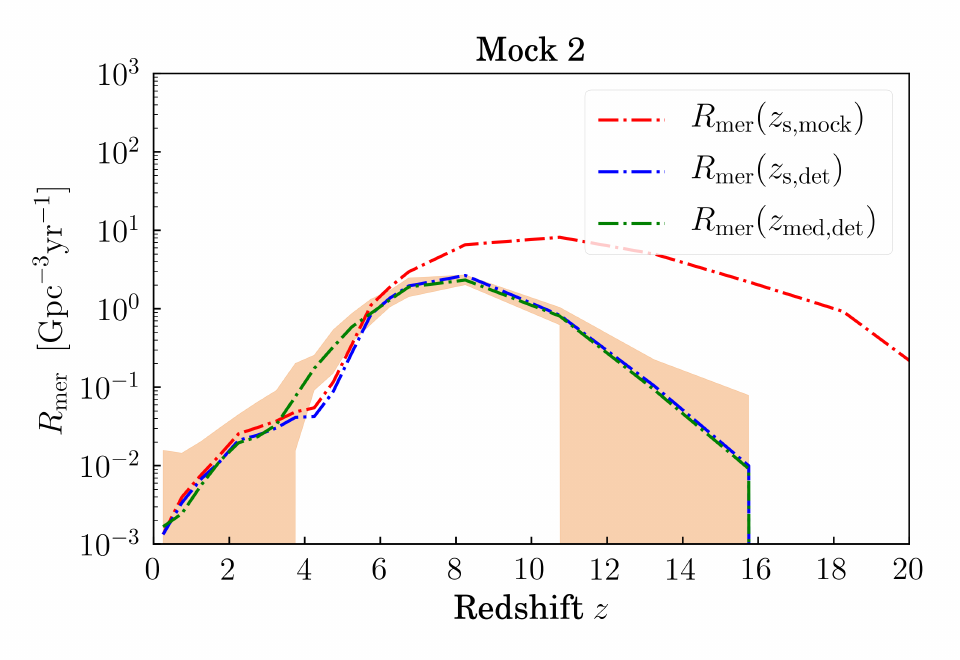}}
\subfloat[\label{fig:100km_rel}]{\includegraphics[width=0.33\textwidth]{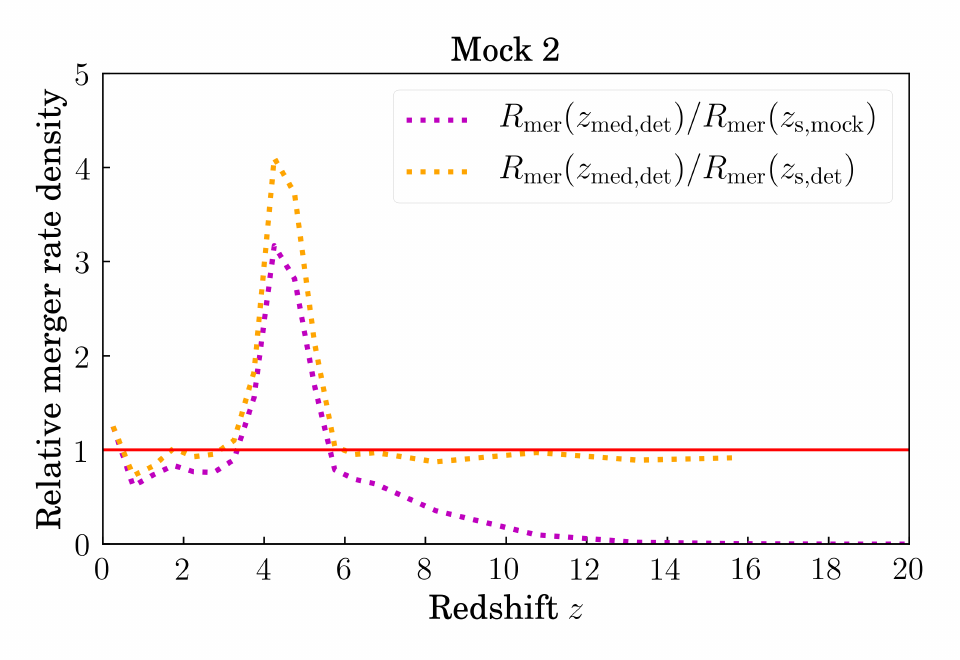}}\subfloat[\label{fig:100km_cummu}]{\includegraphics[width=0.33\textwidth]{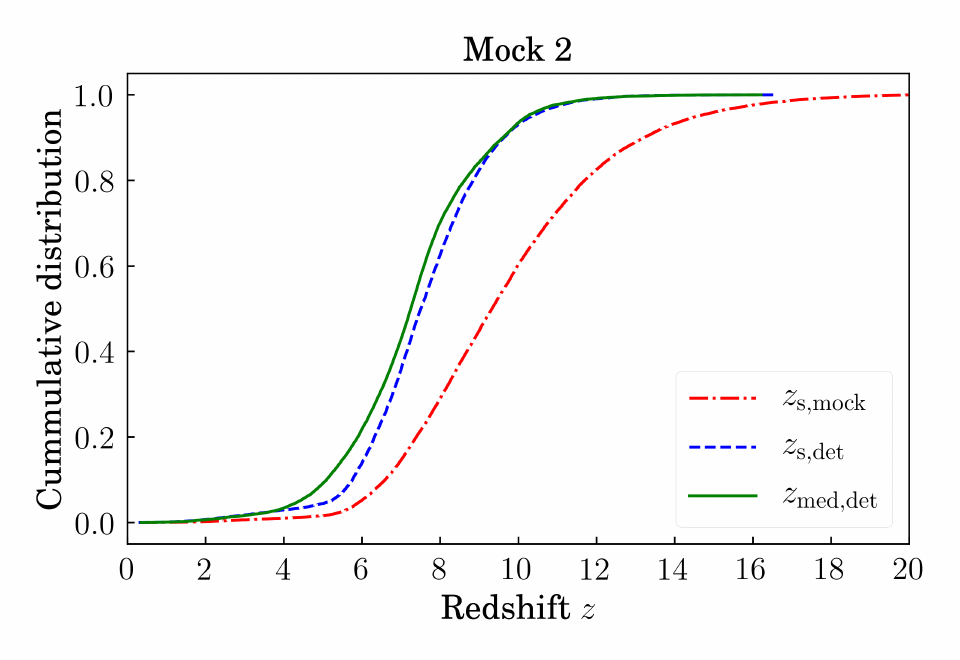}}\\
\subfloat[\label{fig:mixed_rate}]{\includegraphics[width=0.33\textwidth]{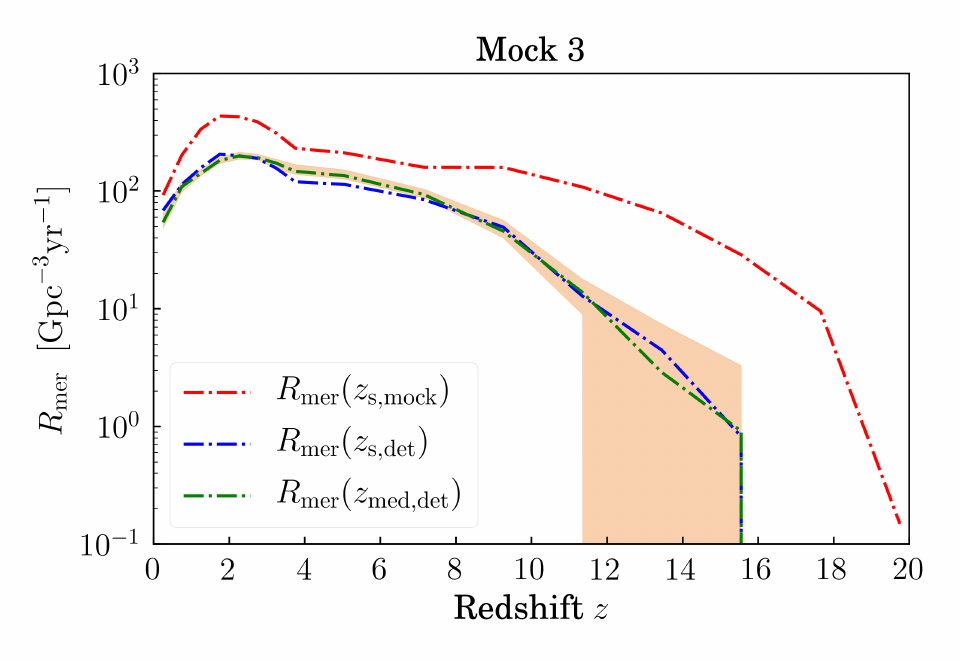}}\subfloat[\label{fig:mixed_rel}]{\includegraphics[width=0.33\textwidth]{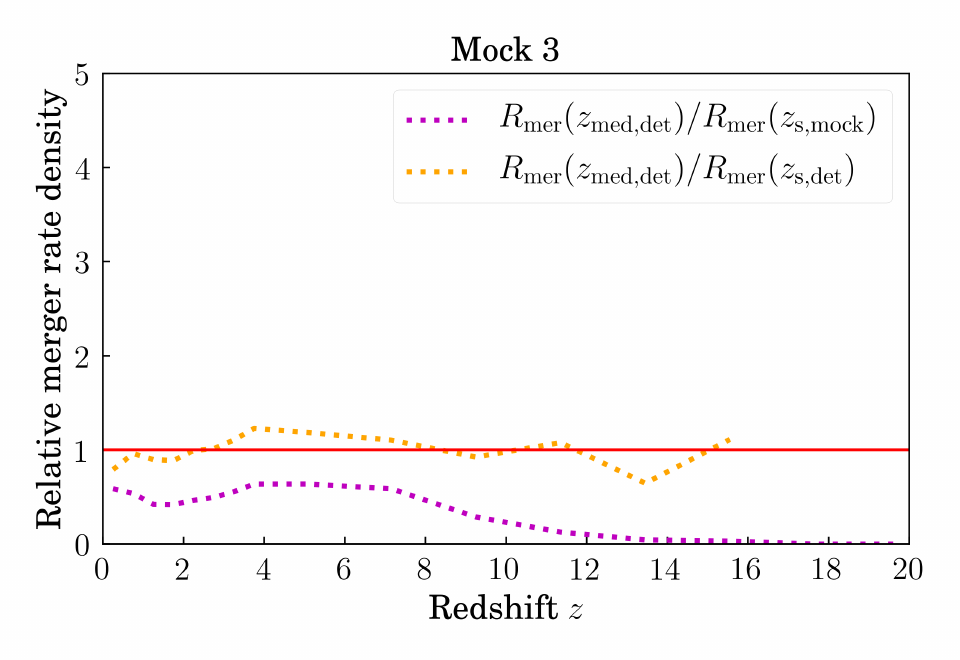}}\subfloat[\label{fig:mixed_cummu}]{\includegraphics[width=0.33\textwidth]{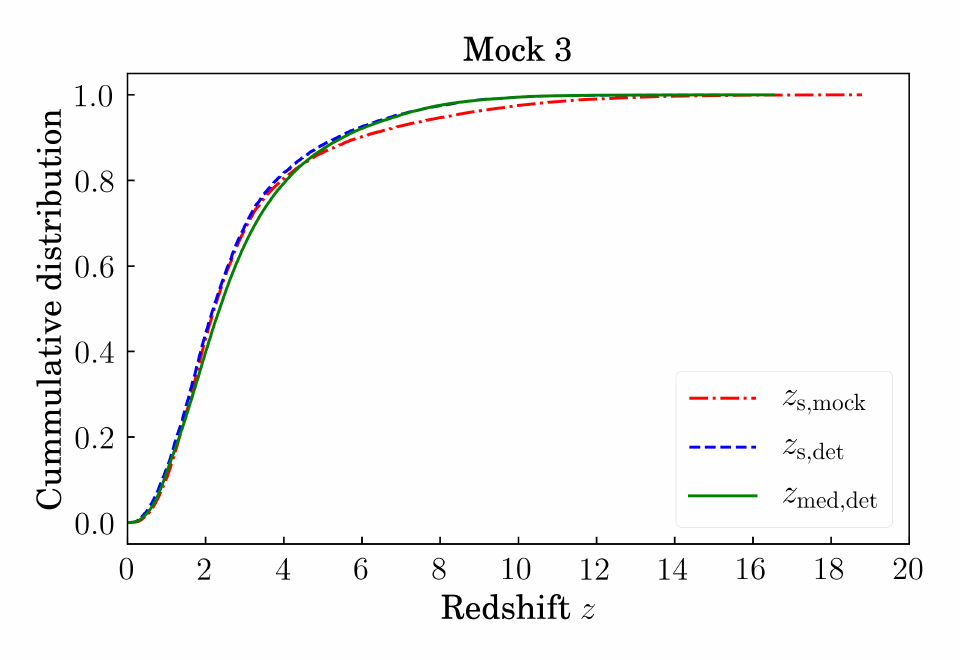}}\\
\caption{Estimate of the merger rate density. Left: Merger rate densities for  Mock 1, Mock 2, and Mock 3, respectively. The shaded region represents Poisson error. Middle: Relative merger rate density. Right: Cumulative probability distribution of the redshifts for Mock 1, Mock 2, and Mock 3. A description of the acronyms is given in Table \ref{tab:acronym}.}
\label{fig:merger_rate}
\end{figure*}

\subsection{Chirp mass and redshift estimates}

The density distribution of the estimated median values with respect to the true values of the parameters for the three populations estimated using the algorithm described above are shown in Figure \ref{fig:injvsmedian}. This figure shows the estimated median values ($\mathcal{M}_{\rm med, det}, z_{\rm med, det}$) with respect to the actual values of the parameters ($\mathcal{M}_{\rm s,det},z_{\rm s, det}$) for each detected compact binary source in three mock populations. The blue contour encloses the 90\% probability region of all the detected sources. The plots in the top panel, namely Figure \ref{fig:50km_chm_med} and \ref{fig:50km_red_med}, can be compared with Figures 3(c,d) of S22 in order to see the effect of using $\mathcal{F}(\rho^i_{21}, \rho^i_{31})$ as an additional prior on $\Lambda$. The estimates of the chirp masses as compared to their true values for the compact binaries in all three populations are shown in Figures \ref{fig:50km_chm_med}, \ref{fig:100km_chm_med}, and \ref{fig:mixed_chm_med}. We see that chirp mass is overestimated for half of the population and underestimate for the other half. Using the current algorithm, the median values of the redshift of 90\% of the detected sources, shown in Figures \ref{fig:50km_red_med}, \ref{fig:100km_red_med}, and \ref{fig:mixed_red_med}, are in agreement with the true values of redshift for Mock 1 and Mock 3, while the redshifts are slightly underestimated for Mock 2. The reason for this it that including $\mathcal{F}(\rho^i_{21}, \rho^i_{31})$ as an additional prior only improves the parameter estimates for those binary sources for which $0.95 \lesssim \rho_{ij} \lesssim 1.05,$ where $\rho_{ij}$ is the ratio of S/N generated in the $i^{th}$ and $j^{th}$ detector of ET. This is explained in more detail in \S \ref{sec:bias_construct} and \S \ref{sec:bias_effect}. We quantify the error in the estimate of the redshift in the following section by calculating the merger rate density for the mock populations.

\section{Estimation of the merger rate density}\label{plan_pop_mer}
In this section, we describe a population-independent method of estimation of the merger rate density as a function of redshift. Using the parameters estimated in the previous section, we calculate the merger rate density for the sources we detect, for a given detection threshold. Then, using the probability distribution of the population parameters of these detected sources, we asses the detection efficiency as a function of redshift. Taking into account this detection efficiency, we then reconstruct the merger rate density for the population of coalescing compact binaries.

\begin{figure*}[!]
\centering
\subfloat[\label{fig:det_eff_1}]{\includegraphics[width=\columnwidth]{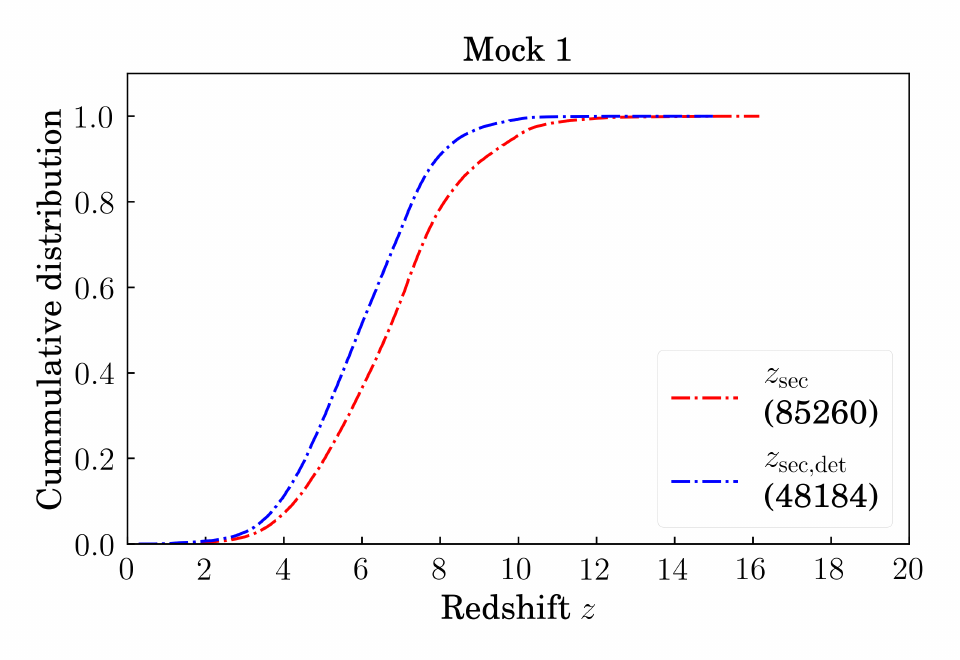}}\subfloat[\label{fig:new_rate_1}]{\includegraphics[width=\columnwidth]{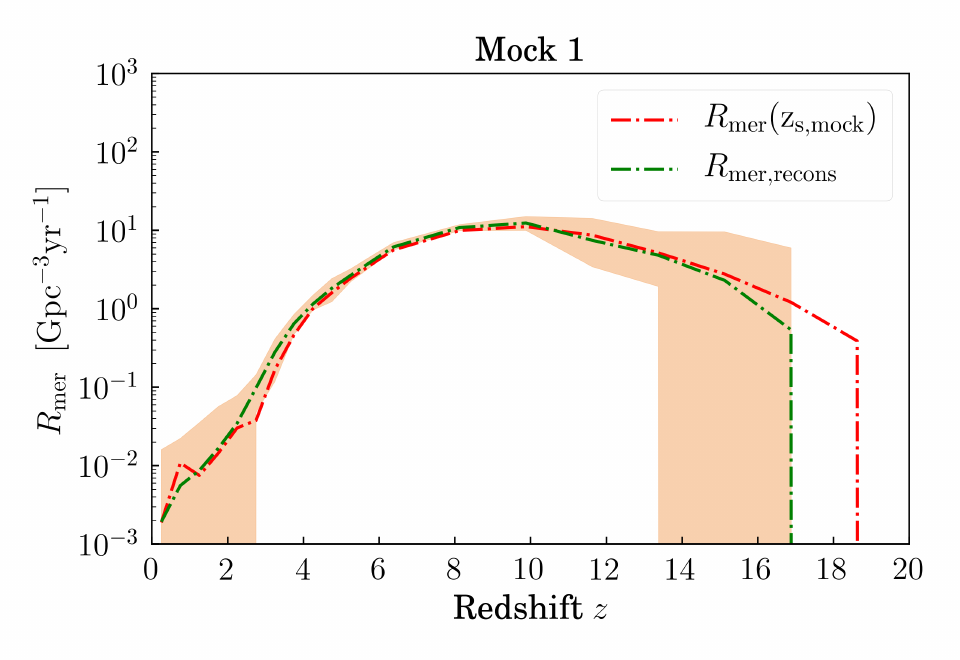}}\\
\subfloat[\label{fig:det_eff_2}]{\includegraphics[width=\columnwidth]{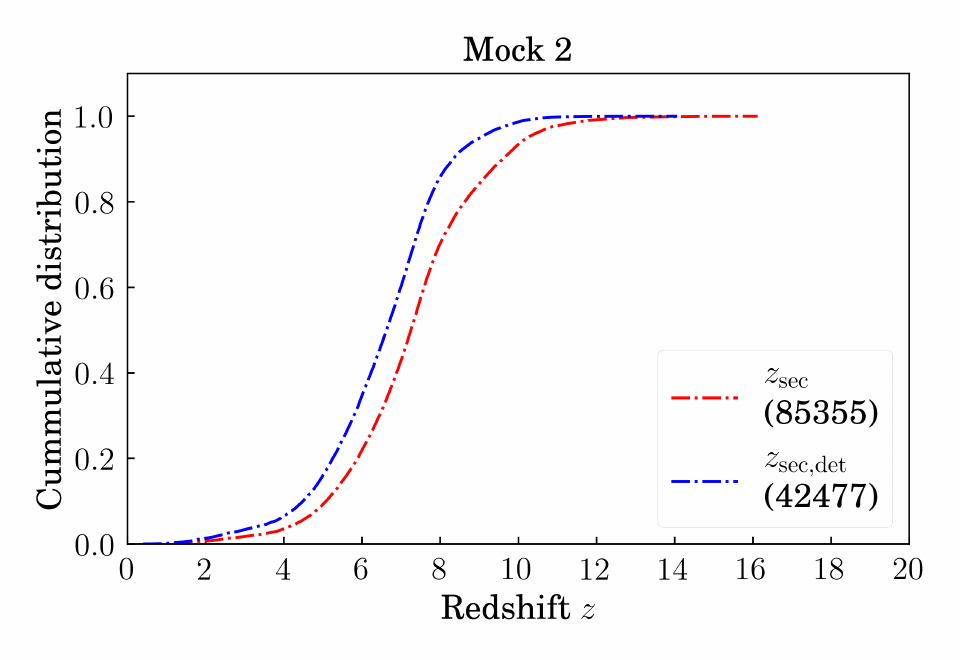}}\subfloat[\label{fig:new_rate_2}]{\includegraphics[width=\columnwidth]{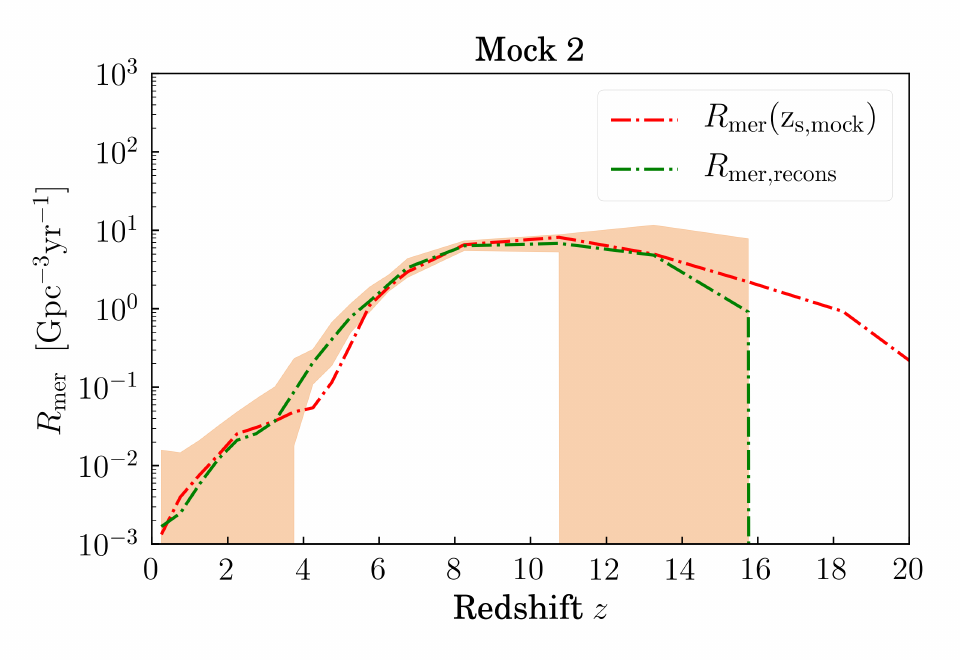}}\\
\subfloat[\label{fig:det_eff_3}]{\includegraphics[width=\columnwidth]{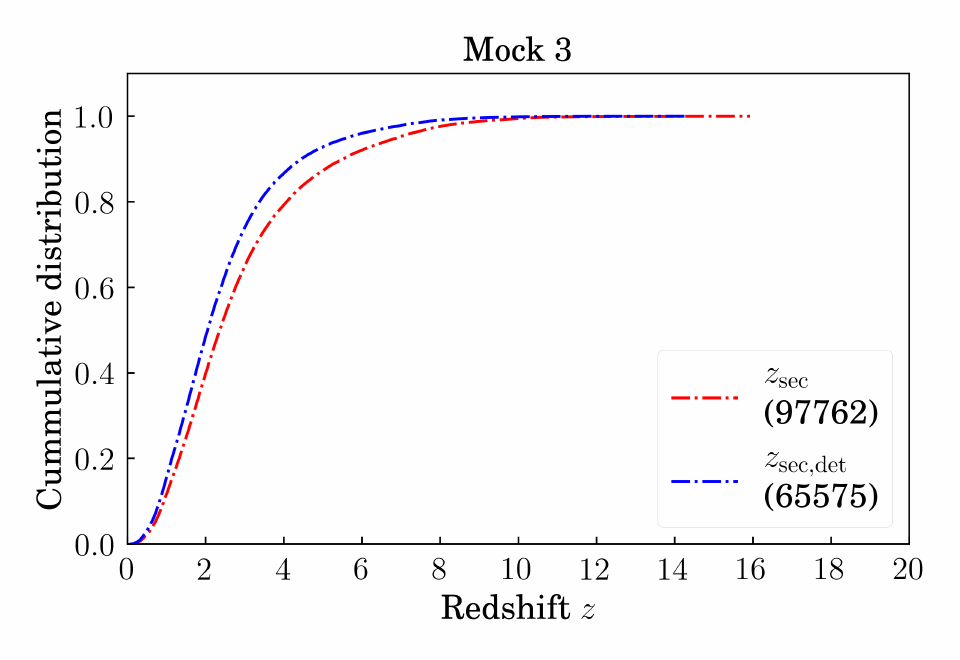}}\subfloat[\label{fig:new_rate_3}]{\includegraphics[width=\columnwidth]{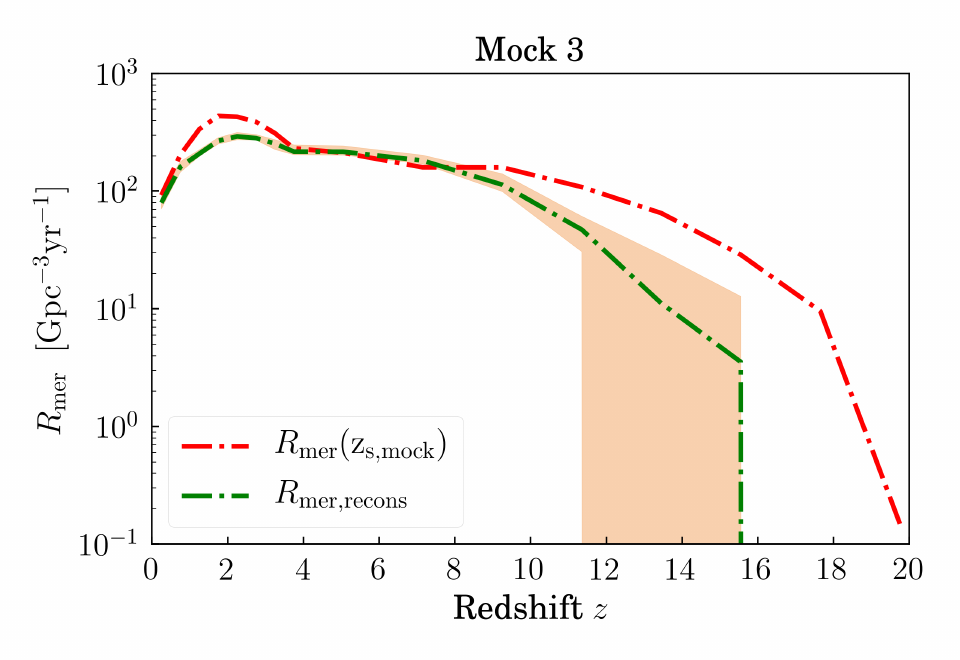}}\\
\caption{Reconstruction of merger rate density. Left: Cumulative probability distribution of the redshift of the secondary population (red) and of the detected sources from these secondary mock populations (blue) for Mock 1, Mock 2, and Mock 3. The number in parentheses is the number of sources in each population. Right: Reconstructed merger rate density for Mock 1, Mock 2, and Mock 3. The shaded region represents the Poisson error.}
\label{fig:det_eff_rate}
\end{figure*}

\subsection{Merger rate density}\label{mer_rate}

For a given population of compact binaries, we simulate $N_{\rm mock}$ number of binaries, of which $N_{\rm det}$ are detected based on the chosen detection threshold. The expected time taken for these binaries in the mock population to merge is 
\begin{equation}
     T_{\rm mock} = \frac{N_{\rm mock}}{N_{\rm yr}} \; \rm{yr},
\end{equation}
where $N_{\rm yr}$ is the number of mergers per year calculated by integrating the merger rate density given by Equation (\ref{Rofi}). The merger rate density $R_{\rm mer}$ for a given population is calculated for the time $T_{\rm mock}$:

\begin{equation}\label{merger_rate_den}
    R_{\rm mer}(z_i, z_{i+1}) = \frac{1+z_{i+1}}{\int^{z_{i+1}}_{z_i} \frac{dV}{dz}dz}\left( \frac{N_{(z_i,z_{i+1})}}{T_{\rm mock}}\right),
\end{equation}
where $N_{(z_i,z_{i+1})}$ is the number of mergers in a redshift bin $[z_i:z_{i+1}]$. 

From Equation (\ref{merger_rate_den}), we obtain three sets of merger rate densities  for a given population: (i) $R_{\rm mer}(z_{\rm s, mock})$, (ii) $R_{\rm mer}(z_{\rm s, det})$, and (iii) $R_{\rm mer}(z_{\rm med, det})$. These merger rate densities are shown in Figures \ref{fig:50km_rate}, \ref{fig:100km_rate}, and \ref{fig:mixed_rate} for Mock 1, Mock 2, and Mock 3, respectively. The shaded region represents the Poisson error. For Mock 1 and Mock 2, we estimated the Poisson error for yearly data sets while for Mock 3 we estimated the Poisson error for monthly data sets.

It can be seen that, for all three populations, $R_{\rm mer}(z_{\rm med, det}) \approx R_{\rm mer}(z_{\rm s, det})$, but $R_{\rm mer}(z_{\rm med, det}) \ll R_{\rm mer}(z_{\rm s, mock})$. This is because most of the merging compact binary systems at higher redshifts do not cross the detection threshold we have chosen. In order to further compare the merger rate densities, we calculated the relative merger rate densities $\frac{R_{\rm mer}(z_{\rm med, det})}{R_{\rm mer}(z_{\rm s, mock})}$ and $\frac{R_{\rm mer}(z_{\rm med, det})}{R_{\rm mer}(z_{\rm s, det})}$. These are shown in Figures \ref{fig:50km_rel}, \ref{fig:100km_rel}, and \ref{fig:mixed_rel}. This calculation shows that $R_{\rm mer}(z_{\rm med, det}) \approx R_{\rm mer}(z_{\rm s, det})$ for Mock 3, while the in cases of Mock 1 and Mock 2, $R_{\rm mer}(z_{\rm med, det})$ is slightly larger than $R_{\rm mer}(z_{\rm s, det})$  in the specific redshift ranges by a factor of $\sim 3$ and $\sim 4,$ respectively.

Figures \ref{fig:50km_cummu}, \ref{fig:100km_cummu}, and \ref{fig:mixed_cummu} show  the cumulative probability distribution of the redshifts for Mock 1, Mock 2, and Mock 3 and we see that $< 1\%$ of the sources in each of the three mock populations are in the redshift range where the relative merger rate densities are $>1.5$. The reason for these few spikes is further explained in \S \ref{sec:bias_effect}. As $R_{\rm mer}(z_{\rm med, det}) \ll R_{\rm mer}(z_{\rm s, mock})$, we calculated the detection efficiency in order to reconstruct the actual merger rate density from the $R_{\rm mer}(z_{\rm med, det})$.

\subsection{Detection efficiency}
Detection of a coalescing compact binary with a gravitation wave detector is ideally defined by the threshold we choose to set for the S/N of such an event.  The higher the S/N threshold value set for detection, the more sources will be left undetected. It is important to note here that, for our analysis, we use the design sensitivity noise curve of ET-D to estimate the S/N and set the threshold for detection, whereas in the case of detections with real noise, one has to take into account the glitches, which will further introduce a selection bias in the detection of events. We neglect the presence of glitches in our analysis. In order to get an accurate estimate of the merger rate density for a population of compact binaries, it is necessary to gauge the number of compact binaries that are not detected as a function of redshift. To this end, we make an assumption that the detected population of the compact binaries truly represents the redshift and chirp mass distribution of the whole population. For a given population, we generate a secondary mock population ---denoted with subscript `sec'--- from the detected sources, assuming that the distributions of the chirp mass and redshift are proportional to $\mathcal{M}_{\rm med,det}$ and $z_{\rm med, det}$, that is, $p(\mathcal{M}_{\rm sec})\propto p(\mathcal{M}_{\rm med,det})$ and $p(z_{\rm sec}) \propto p(z_{\rm med,det})$. We assume that the mass ratio $q_{\rm sec}$ is uniformly distributed in the range [0,1] with a constraint on total mass $M_{\rm sec}$ such that $(M_{\rm med,det})_{\rm min} \leq M_{\rm sec} \leq (M_{\rm med,det})_{\rm max}$, where $M_{\rm sec}$ is defined as:

\begin{equation} \label{totalmass}
   M_{\rm sec} = \mathcal{M}_{\rm sec} \left[\frac{q_{\rm sec}}{(1+q_{\rm sec})^2} \right]^{-3/5} .
\end{equation}

To each compact binary of this generated  secondary population, we assign random values to the four angular parameters: the right ascension $\alpha$, the angle of declination $\delta$, the polarisation angle $\psi,$ and the inclination angle $\iota$ of the binary with respect to the direction of observation. The values of $\cos\delta, \alpha/ \pi$, $\cos \iota,$ and $\psi/ \pi$ are chosen to be uncorrelated and distributed uniformly over the range $[-1,1]$. We now use this secondary population to estimate the detection efficiency. Given the detection threshold we chose, the detection efficiency $\mathcal{D}$ as a function of redshift is defined as

\begin{equation}\label{det_eff}
    \mathcal{D}(z_i, z_{i+1}) = \left[\frac{N_{\rm sec, det}}{N_{\rm sec}}\right]_{(z_i,z_{i+1})}
,\end{equation}
where $[N_{\rm sec}]_{(z_i,z_{i+1})}$ is the number of mergers in the secondary mock population in the redshift bin $(z_i,z_{i+1})$ and $[N_{\rm sec, det}]_{(z_i,z_{i+1})}$ is the mergers in this bin that crossed the detection threshold. 

The cumulative probability distributions of the secondary mock populations for each of the three mock populations, that is, Mock 1, Mock 2, and Mock 3, are shown in the left panel of Figure \ref{fig:det_eff_rate} in red, while the cumulative probability distributions of the detected sources from these secondary mock populations are shown in blue. For a given mock population, we use the ratio of the gradient of these curves ---taking into account the number of sources in each set of populations---  to quantify the detection efficiency. 

\subsection{Reconstructed merger rate density}
Now we can reconstruct the merger rate density taking into account the detection efficiency calculated in the previous section. The reconstructed merger rate density $R_{\rm mer, recon}$ is then given as

\begin{equation}\label{recons_mer}
    R_{\rm mer, recon}(z_i, z_{i+1}) = \left[\frac{R_{\rm mer}(z_{\rm med, det})}{\mathcal{D}}\right]_{(z_i, z_{i+1})}
.\end{equation}
The reconstructed merger rate density is calculated for the three mock populations using Equation (\ref{det_eff}) in (\ref{recons_mer}). The reconstructed merger rate densities for Mock 1, Mock 2, and Mock 3 are shown in the right panel Figure \ref{fig:det_eff_rate}. The red lines are the true merger rate density, while the green lines show the merger rate density calculated using Equation (\ref{recons_mer}). The shaded region represents the Poisson error. It can be seen that the merger rate density is reconstructed accurately up to redshift $z\sim 15$ for the Mock 1 population, and up to redshift $z\sim 14$  and Mock 2. 

In the case of Mock 3 binaries, the reconstructed merger rate density at $z\sim 2$ is a factor of $\sim 1.3$ smaller that the true merger rate density. This is so because we calculate the detection efficiency by generating a secondary mock population, assuming that the probability distribution of the chirp mass and redshift is proportional to that of $\mathcal{M}_{\rm med,det}$ and $z_{\rm med, det}$, respectively. However, in the case of Mock 3 binaries, the merger rate density at $z \lesssim 2$ is dominated by low-mass binaries (see Fig. 12 in \citet{2020A&A...636A.104B}), and as we do not detect the bulk of these low-mass binaries given the chosen detection threshold (see Fig. in S22 ), the mass distribution for these objects is not truly represented in the secondary population constructed for Mock 3. The reconstructed merger rate density for Mock 3 is also lower than the true value for $z>8$. As seen in Figure \ref{fig:mixed_chm_med}, Pop III binaries constitute a small percentage of Mock 3 binaries, and so they are also under-represented in the secondary population. In order to have an accurate estimate of the detection efficiency in a mixed population of binaries where the merger rate densities of the two individual populations differ by a large factor, it is essential to have a larger observational data set in order for the underlying populations to be accurately represented.

\begin{figure*}
\centering
\subfloat[\label{fig:sfr_50_ideal}]{\includegraphics[scale=0.45,right]{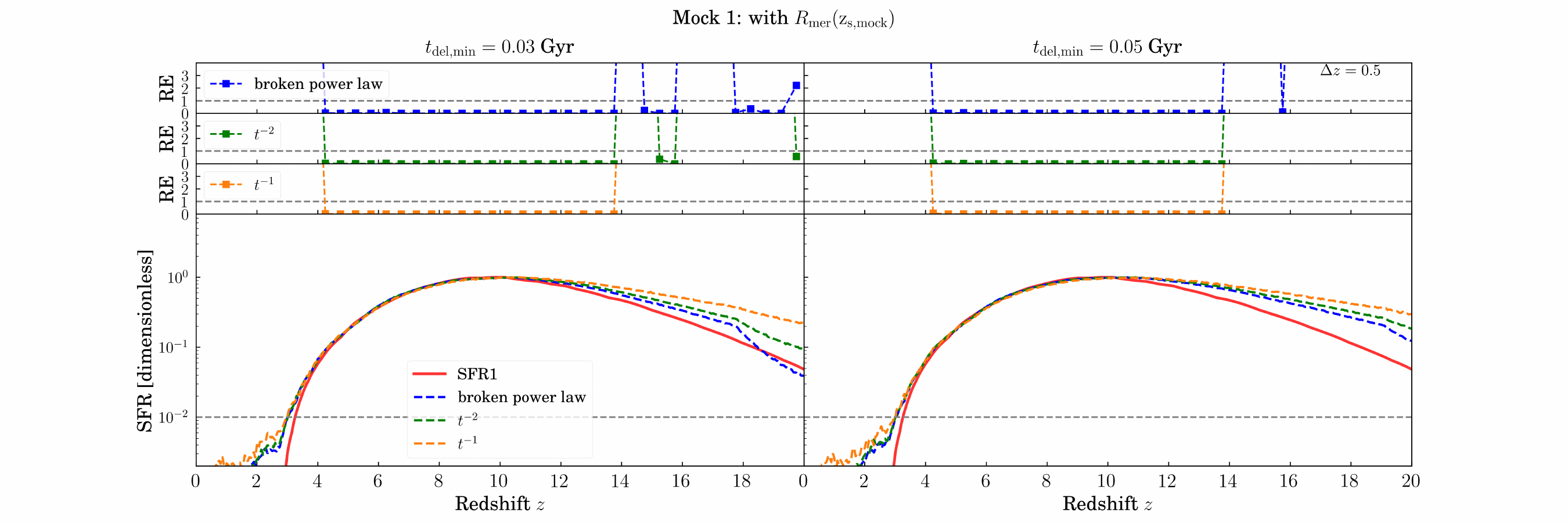}}\\
\subfloat[\label{fig:sfr_100_ideal}]{\includegraphics[scale=0.45,right]{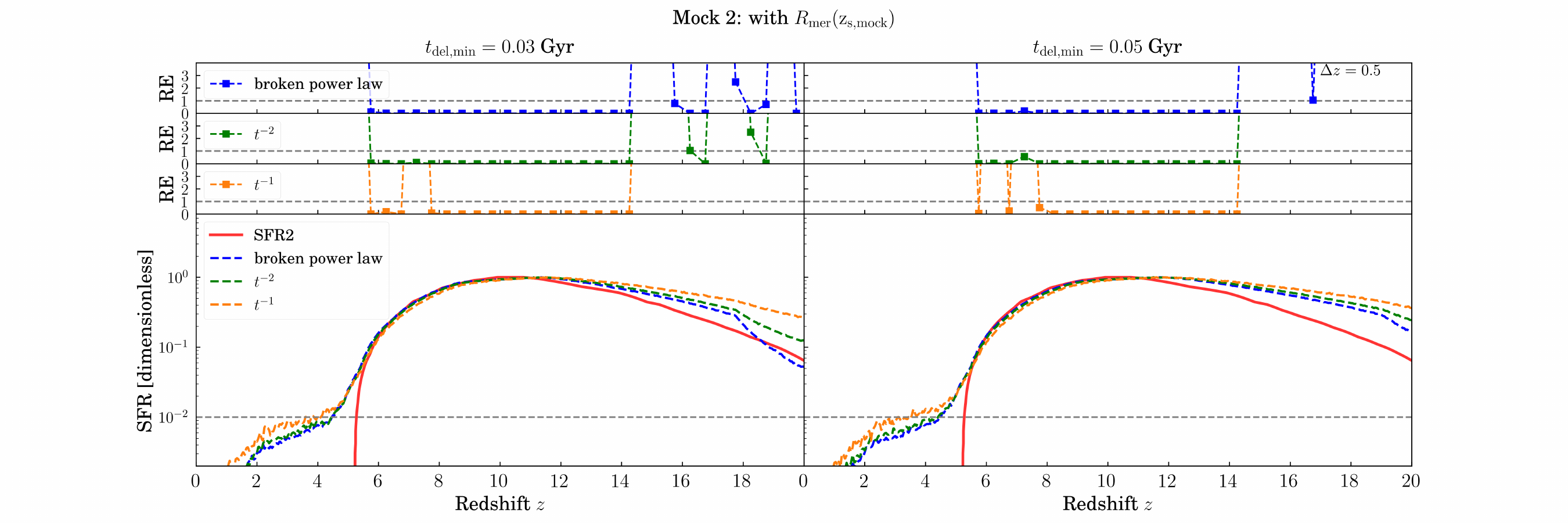}}\\
\subfloat[\label{fig:sfr_mixed_ideal}]{\includegraphics[scale=0.45,right]{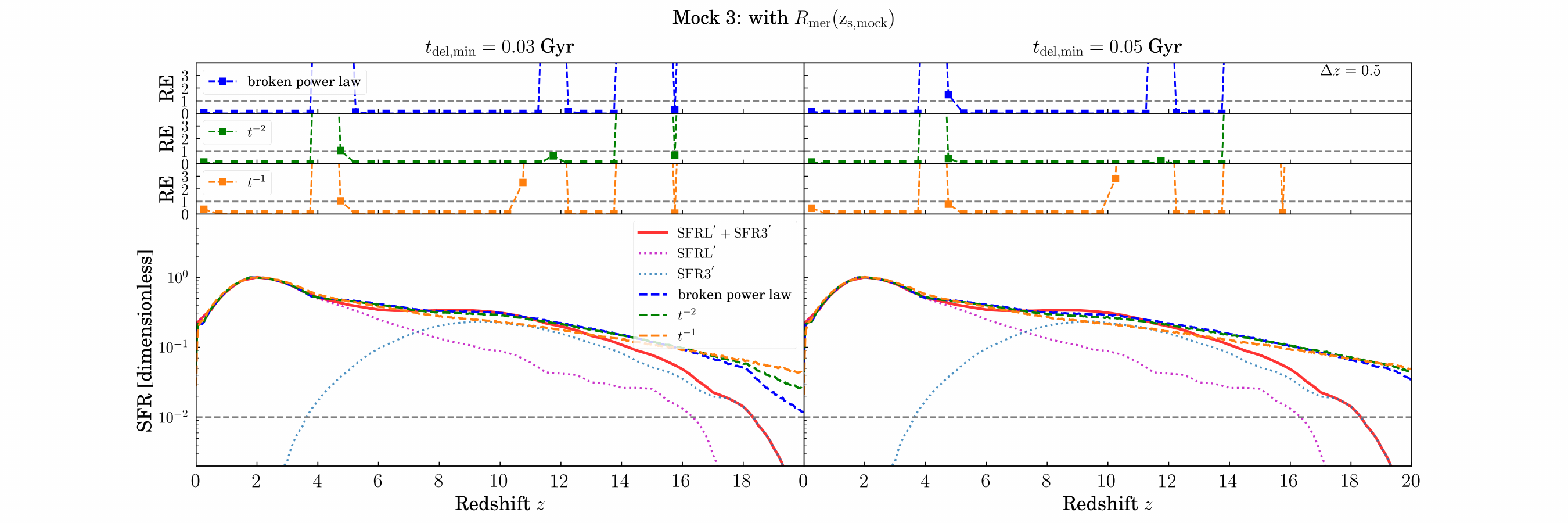}}\\
\caption{Reconstructed SFRs for the three populations, assuming that the merger rate density is highly accurate. The maximum value of the SFR has been normalised to 1. The left and right panels show the reconstruction assuming that $t_{\rm del, min} = 0.03$ and $0.05$ Gyr, respectively. The top three panels in all three plots show RE or the KL divergence for a redshift bin of  $\Delta z = 0.5$ in width. The red curve denotes the true SFR. We note that for the Mock 3 population, we estimate the merger-rate-weighted SFR.}
\label{fig:sfr_ideal}
\end{figure*}

\begin{figure*}
\centering
\subfloat[\label{fig:sfr_50}]{\includegraphics[scale=0.45,right]{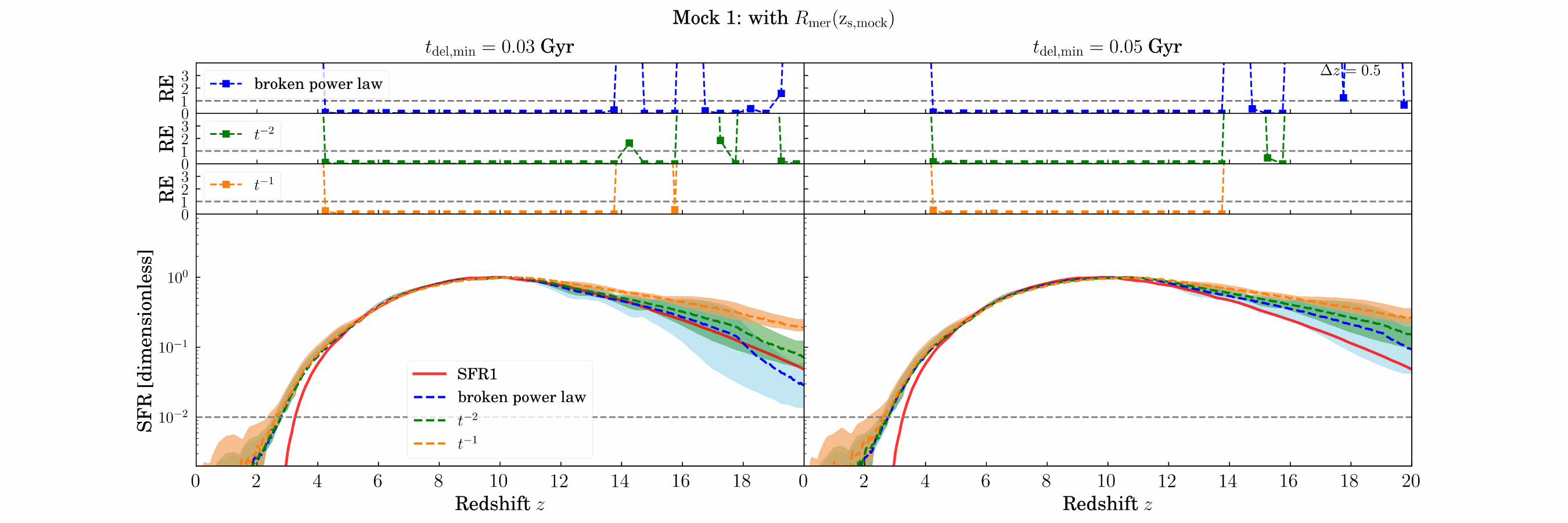}}\\
\subfloat[\label{fig:sfr_100}]{\includegraphics[scale=0.45,right]{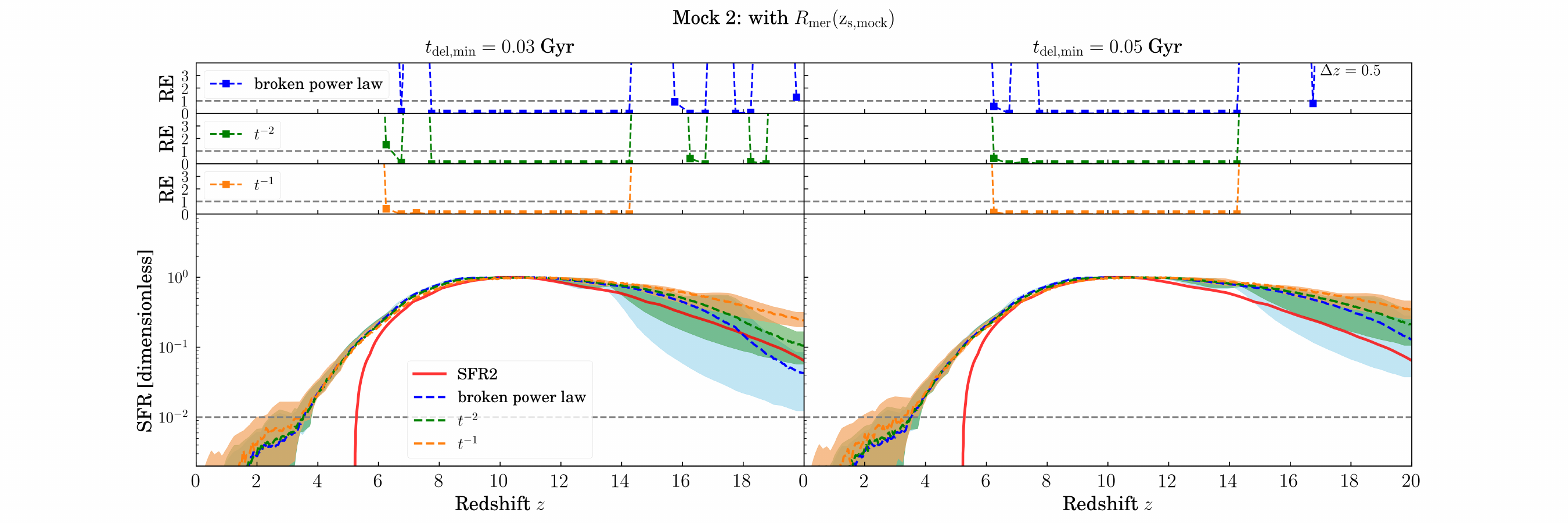}}\\
\subfloat[\label{fig:sfr_mixed}]{\includegraphics[scale=0.45,right]{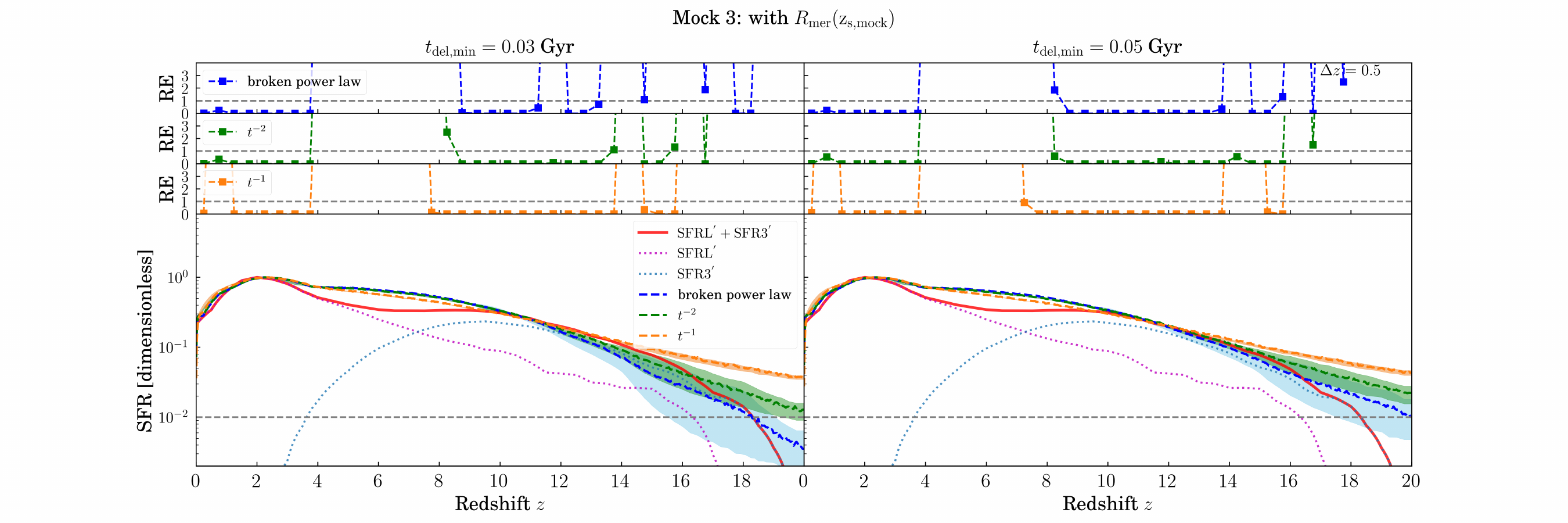}}\\
\caption{Reconstructed SFRs for the three populations with the reconstructed merger rate density $R_{\rm mer, recon}$. The maximum value of the SFR has been normalised to 1. The left and right panels show the reconstruction assuming that $t_{\rm del, min} = 0.03$ and $0.05$ Gyr, respectively. The top three panels in all three plots show RE or the KL divergence for a redshift bin of $\Delta z = 0.5$ in width. The red curve denotes the true SFR. We note that for the Mock 3 population, we estimate the merger-rate-weighted SFR.}
\label{fig:sfr_final}
\end{figure*}

\section{Star formation rates}\label{sfr_rate}

We now proceed to estimate the SFR using our estimate of the merger rate densities. We calculate the SFR using $R_{\rm mer, recon}$, assuming three different time-delay $t_{\rm del}$ distributions: (i) a broken power law , (ii) $t^{-1}$, and (iii) $t^{-2}$, where $t_{\rm del, min} < t < t_{\rm H}$ Gyr. The broken power-law distribution is assumed to be:

\begin{equation}
p(t_{\rm del}) \propto \left\{
\begin{array}{ll}
t^{-3} & {\rm if \ \ \ } t < 1 \rm{Gyr}\\
t^{-2} & {\rm if \ \ \ } t \geqslant 1 \rm{Gyr}.
\end{array}  \right.
\end{equation}
The broken power law time-delay distribution is motivated by the fit to the actual time-delay distribution as shown in Figure 11 of \citet{2017MNRAS.471.4702B} for FS1 compact binaries. The time-delay distribution $t^{-1}$ is based on the best fit for the Pop I+II compact binaries (M30B time-delay data is available on the StarTrack\footnote{http://www.syntheticuniverse.org/} website). The SFR estimate at a given $z$ is the sum over contributions from each  binary `$b$', averaged over its delay time:

\begin{equation}\label{sfr}
    {\rm{SFR}}(z) =  M_{\rm sim}\sum_b \left(\int^{t_{\rm H}}_{t_{\rm del, min}} R^b_{\rm mer, recon}\frac{dz_{\rm mer}}{dt_{\rm del}} p(t_{\rm del}) dt_{\rm del} \right ),
\end{equation}
for $ z^b_{\rm ini} \equiv z(t^b_{\rm mer}-t^b_{\rm del})$. $M_{\rm sim}$ is a constant, and is dependent on the formation scenario. Thus, $M_{\rm sim}$ provides a normalisation for a formation channel. 
We assume two different values for minimum time delay: $t_{\rm del, min} = 0.03$ and $0.05$ Gyr. We denote the Hubble time  $t_{\rm H}$ . In this analysis, we normalise the SFR to one, because the goal is to estimate the functional form of the SFR with redshift. We therefore remove any dependence on the formation channel by removing the dependence on $M_{\rm sim}$ in the case of Mock 1 and Mock 2 populations, where all the binaries are assumed to be from a single population. In the case of Mock 3, where we have a mixed population, our algorithm estimates a merger-rate-weighted SFR as a function of redshift instead of the true sum of two individual SFRs. We denote this merger-rate-weighted SFR as $\rm SFR^{'}$.

It should also be noted that while the three mock populations constructed in this analysis are built with the time-delay distribution that was the output of the StarTrack code and so encode the uncertainties for the formation channel parameters, we reconstruct the SFR assuming the three different time-delay $t_{\rm del}$ distributions described above. We compare the estimated values of the SFRs with the true value of the SFR by calculating the  relative entropy (RE), otherwise known as the Kullback–Leibler divergence ($D_{KL}$) \citep{cover1999elements, kullback1951information,2014arXiv1404.2000S}. RE, or $D_{KL}$, quantifies how close a given probability distribution $p = {p_i}$ is to a given model distribution $q = {q_i}$. It can also be said that the $D_{KL}(p||q)$ is a measure of the inaccuracy of the assumption that the distribution is $q$ when the true distribution is $p$:

\begin{equation}
   {\rm RE} \equiv D_{KL}(p||q) = \sum_{i} p_i \log \left( \frac{p_i}{q_i}\right).
\end{equation}
RE is always non-negative, and it is zero only under the condition that $p=q$ \citep{cover1999elements}. We assume that for RE > 1, the distributions cannot be considered to be similar.

To understand the errors in the reconstructed SFR due to the different time-delay distributions and due to the error in the estimated merger rate density, we proceed as follows. As a first step, we estimate the SFR assuming that merger rate densities for each of the three populations $R_{\rm mer}(z_{\rm s, mock})$ are known with high accuracy, so that we can see the effect of only the variation of true time-delay distribution ---as compared to the assumed time-delay distributions--- on the reconstructed SFR. Then, as a second step, we reconstruct the SFR with $R_{\rm mer, recon}$ for each of the three mock populations in order to estimate the errors on the SFR due to both the error on the estimated merger rate density and the variation of the true time-delay distribution in comparison to the assumed time-delay distribution.

Figure \ref{fig:sfr_ideal} shows the SFR reconstructed for Mock 1 (top), Mock 2 (middle), and Mock 3 (bottom), assuming an accurate merger rate density with redshift $R_{\rm mer}(z_{\rm s, mock})$. In each of these three figures, the left panel shows the reconstruction assuming that $t_{\rm del, min} = 0.03$ Gyr, and the right panel shows the results for $t_{\rm del, min} = 0.05$ Gyr. The top three panels in all three plots show the RE for a redshift bin of $\Delta z = 0.5$ in  width. 

For each of the three mock populations, the reconstruction of SFR as a function of redshift is independent of the assumed time-delay distribution throughout almost the entire redshift range. The comparison of the reconstructed SFR with the true SFR using KL divergence shows that ${\rm RE} < 1$ from nearly the termination redshift up to $z \sim 14$. In a few redshift bins, ${\rm RE} > 1$ due to the fact that the assumed time-delay distributions are not exact representations of the true time-delay distributions of these mock populations. It should be noted that the assumed time-delay distributions in this analysis do not correctly model the longer time-delay distribution $t_{\rm del} >8$ Gyr. This is evident in the case of Mock 2 in Figure \ref{fig:sfr_100_ideal}. While it is crucial to have a better model for long time-delay distribution in order to accurately estimate the termination redshift of the SFR, accurate information about the minimum time delay $t_{\rm del, min}$ is essential in order to estimate the SFR at redshifts beyond $z \sim 14$. For Mock 1 and Mock 2, the true values of the termination redshift (which we define as the redshift where the SFR is 1\% of its peak value) are $\sim 3.2$ and  $\sim 5.3$. As seen from Figure \ref{fig:sfr_ideal}, the errors on the estimates of the termination redshifts due to incorrect modelling of the tail of the time-delay distribution are $\Delta z/z \lesssim 7\%$ and $\Delta z/z \lesssim 32\%$ for Mock 1 and Mock 2, respectively.

Regarding the reconstruction of the SFR for Mock 3 shown in Figure       \ref{fig:sfr_mixed_ideal}, an important point to note is that the estimated SFR is a rate-weighted SFR, and so $\rm SFR^{'} \neq SFRL+ SFR3$. The reconstructed SFR clearly shows the presence of two different populations, with one formation peaking at $z\sim 2$ and another peaking at $z\sim 10$.
As in the case of Mock 1 and Mock 2, the RE values in the case of Mock 3 show that the reconstruction is independent of the time-delay distributions up to $z \sim 14$ and the accuracy of the reconstruction of SFR strongly depends on the time-delay distribution only at higher redshifts of $z\gtrsim 14$. We further verify this conclusion by assuming a few more extreme time-delay distributions. A discussion about the reconstruction of the SFR with these extreme time-delay distributions is presented in Sect \ref{extreme_delay} of the Appendix.

We now proceed to reconstruct the SFR with $R_{\rm mer, recon}$ for each of the three mock populations. Figure \ref{fig:sfr_final} shows the estimated SFRs for Mock 1, Mock 2, and Mock 3. The true SFR is shown in red for each of these three populations. The left panel in each figure shows the reconstructed SFR assuming that $t_{\rm del, min} = 0.03$ Gyr, and the right panel shows the results for $t_{\rm del, min} = 0.05$ Gyr. The shaded region represents Poisson error. The RE comparing the estimated values of the SFRs with the true value of the SFR in redshift bins of $\Delta z = 0.5$ in  width is shown in the top three panels for each of the three populations.

As before, we see that  for each of the three mock populations, the reconstruction of the SFR as a function of redshift is independent of the assumed time-delay distribution and the comparison of the reconstructed SFR with the true SFR using KL divergence shows that RE < 1 from nearly the termination redshift up to $z \sim 14$. The combined error on the termination redshift is due to the error from the improper modelling of the long time-delay distributions and the error on the estimates of the merger rate density as a function of redshift.

The top three panels in Figure \ref{fig:sfr_50}, Figure \ref{fig:sfr_100}, and Figure \ref{fig:sfr_mixed} show that RE< 1 for $4 \lesssim z \lesssim 14$ for Mock 1, that RE< 1 for $6 \lesssim z \lesssim 14$ for Mock 2, and that RE< 1  for $4 \lesssim z \lesssim 14$ for Mock 3, respectively, irrespective of the assumed time-delay distribution. The errors on the estimate of the termination redshift for Mock 1 and Mock 2 are $\Delta z/z \lesssim 22\%$ and $\Delta z/z \lesssim 37\%,$ respectively.

For Mock 3, which is a mixed population of Pop I+II+III, we show the reconstructed SFR in Figure \ref{fig:sfr_mixed}. As seen in Figure \ref{fig:new_rate_3}, the $R_{\rm mer, recon} < R_{\rm mer}(z_{\rm s, mock})$ for $z\sim 2$ and $z>8$ because of the underestimation of the detection efficiency. As a result of this, RE < 1 up to $z \lesssim 14,$ except in the range $4 \lesssim z \lesssim 8$. The reconstructed SFR for this mixed population clearly shows the presence of the peaks of the two different populations, with one formation peaking at $z \sim 2$ and another peaking at $z \sim 10$. This estimate can be further improved with a larger observational data set, because this would provide a better representation of the underlying populations.

As we only use the inspiral part  in this analysis to estimate the parameters of the compact binaries, as done previously in SB1, SB2, and S22, the S/N generated in the ET detectors is underestimated. The estimates on the parameters are therefore conservative, and so the errors estimated on the reconstructed SFR are the upper bounds.

\section{Conclusion}\label{conc}

In this paper, we used an updated version of the  SB2 algorithm to estimate parameters such as chirp mass, redshift, total mass, and mass ratio for compact binaries. We reconstructed the merger rate density and SFR for three mock population models using single ET and assuming a triangular configuration and ET-D design sensitivity. We constructed the mock populations for compact binaries originating in stars from Population (Pop) I+II and Pop III,  assuming different SFRs and realistic time-delay distributions.  

For a given population, as a first step, we estimated the chirp mass, redshift, and total mass of each detected compact binary. We then estimated the detection efficiency for each population and thus reconstructed the merger rate density taking into account the fraction of binaries that do not cross the detection threshold. We then reconstructed the SFRs using the estimated merger rate density assuming three different functional forms for $t_{\rm del}$ and two different values of $t_{\rm del, min}$. The variable names mentioned in the text are summarised in Table \ref{tab:acronym}.

For Mock 1 and Mock 2, the true values of termination redshifts are $z \sim 3.2$ and  $z \sim 5.3$, and the errors on the estimation of the these termination redshifts due to incorrect modelling of tail of the time-delay distribution are $\Delta z/z \lesssim 7\%$ and $\Delta z/z \lesssim 32\%$ for Mock 1 and Mock 2, respectively. Taking into account the error on the estimates of the merger rate density in addition to the incorrect modelling of the tail of time-delay distribution, we estimate that the errors on the termination redshift are $\Delta z/z \lesssim 22\%$ and $\Delta z/z \lesssim 37\%$ for Mock 1 and Mock 2, respectively. We conclude that the farther the true termination redshift is, the larger the error on the estimate will be given the inaccurate modelling of the tail of the time-delay distribution.

For Mock 3, which is a mixed population of Pop I+II+III, the reconstructed merger rate density at $z\sim 2$ is a factor of $\sim 1.3$ smaller that the true merger rate density, because the detection efficiency calculated by generating a secondary mock population assumes that the probability distributions of chirp mass and redshift of the underlying population are represented by the distributions of the chirp mass and redshift of the detected population, respectively. However, in the case of Mock 3 binaries, the merger rate density at $z \lesssim 2$ is dominated by low-mass binaries, and given that we do not detect the bulk of these objects, with the chosen detection threshold, the mass distribution for these low-mass binaries is not truly represented in the detected population for Mock 3. The reconstructed merger rate density for Mock 3 is also lower than the true value for $z>8$. As Pop III binaries constitute a small percentage of Mock 3 binaries, they are under-represented in the secondary population. This estimate can be further improved with a larger observational data set, because it will provide a better representation of the underlying populations and thus improve our estimate of the detection efficiency. 

In conclusion, we provide a method to reconstruct the functional form of the SFR for populations of compact binaries with ET. The SFR as a function of redshift is accurately reconstructed up to redshift  $z \sim 14$. For all three of our mock populations, we show that the reconstruction of SFR is independent of the time-delay distributions up to $z \sim 14$. The accuracy of the reconstruction of the SFR beyond  $z \sim 14$ strongly depends on the minimum value of the time delay $t_{\rm del, min}$. The assumed time-delay distributions in this analysis do not correctly model the longer time-delay distribution $t_{\rm del} >8$ Gyr. While we accurately reconstruct the SFR as a function of redshift for the bulk of each mock population, a better model for the long time-delay distribution is needed in order to accurately estimate the termination redshift of the SFR. We therefore constrained the peak of the SFR as a function of redshift, and show that ET as a single instrument can distinguish the termination redshifts of different SFRs if they have a true separation of at least $\Delta z \sim 2$.

While we used only one population-evolution model for Pop III (FS1) and Pop I+II (M30B), using different population models will not effect the recovery of the peak of the SFRs unless the peak is beyond $z\sim 14$; beyond this redshift, we need accurate time-delay distributions to estimate the SFR. Using different population models will also not effect the recovery of the termination redshift because the error comes only from the inaccurate modelling of the tail of the time-delay distribution. For any given population, if the SFR terminates at some redshift of $z\sim 6$ for example and if we still detect the binaries from this population at $z\sim 0-1$, this means these are the binaries with long time delays. The farther the true termination redshift is, the greater the probability that the binaries we detect today will be those with long time delays. This is the reason for the larger error on the estimate of termination redshift for Mock 2 as compared to Mock 1.

\begin{acknowledgements}
We thank the anonymous referee for very helpful comments and suggestions.  NS is supported by the "Agence Nationale de la Recherche", grant n. ANR-19-CE31-0005-01 (PI: F. Calore) and is thankful to Astronomical Observatory, University of Warsaw for providing access to computing resources. TB and MC are supported by the grant "AstroCeNT: Particle Astrophysics Science and Technology Centre" (MAB/2018/7) carried out within the International Research Agendas programme of the Foundation for Polish Science (FNP) financed by the European Union under the European Regional Development Fund. TB acknowledges support of NCN through the Harmonia grant UMO-2017/26/M/ST9/00978. This document has been assigned LAPTh document number LAPTH-016/23.
\end{acknowledgements}

\bibliography{reference}
\bibliographystyle{aa}

\begin{appendix}

\section{Construction of $\mathcal{F}$ }\label{sec:bias_construct}

\begin{figure*}[!]
\centering
\subfloat[\label{fig:ratio_lambda}]{\includegraphics[scale = 0.6]{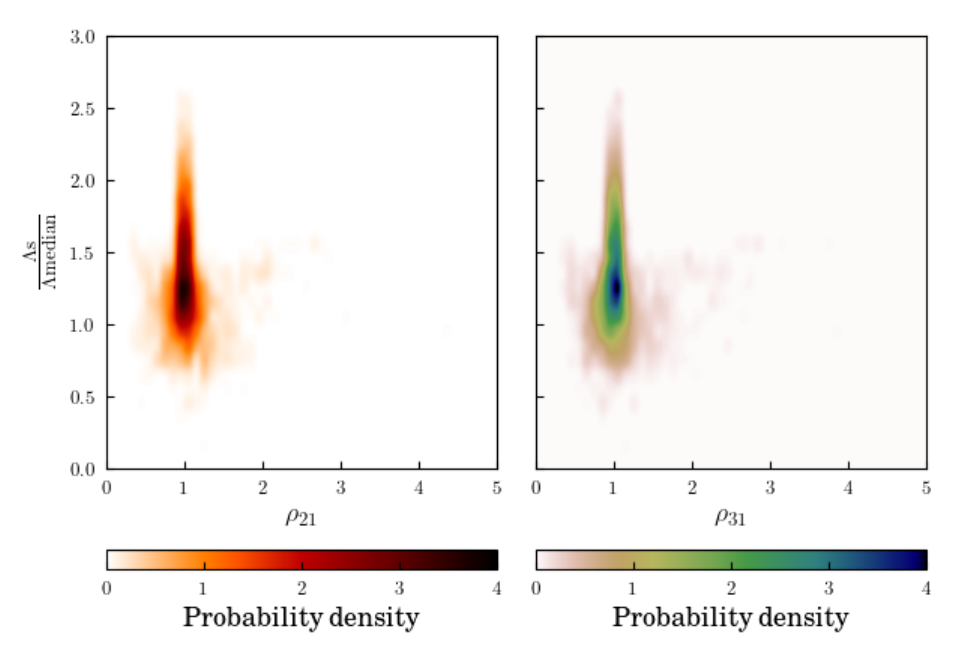}}\\
\subfloat[\label{fig:fit_3d}]{\includegraphics[scale = 0.65]{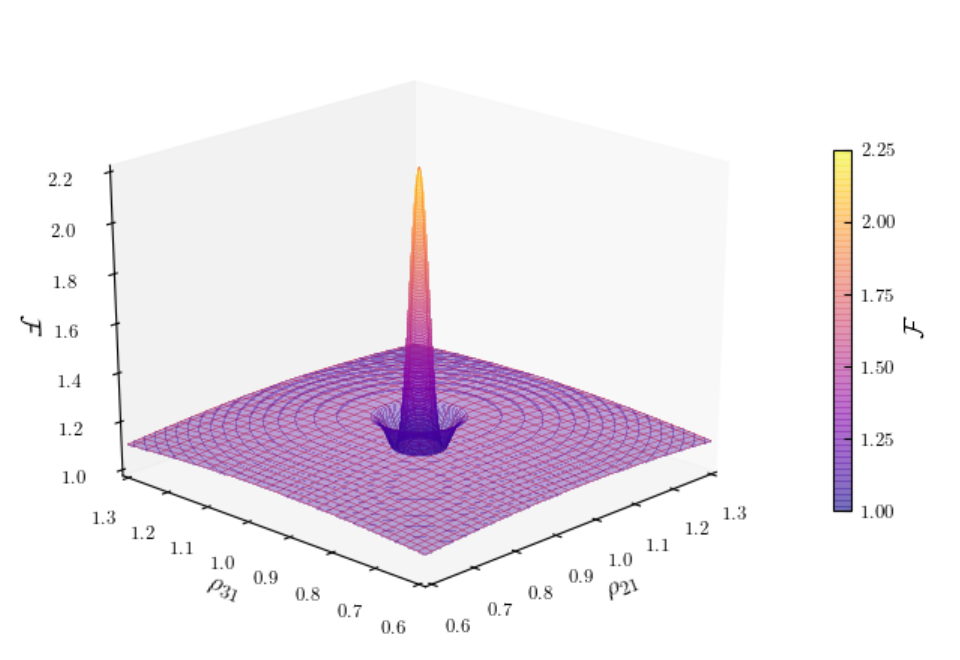}}\\
\subfloat[\label{fig:fit_2d}]{\includegraphics[scale = 0.6]{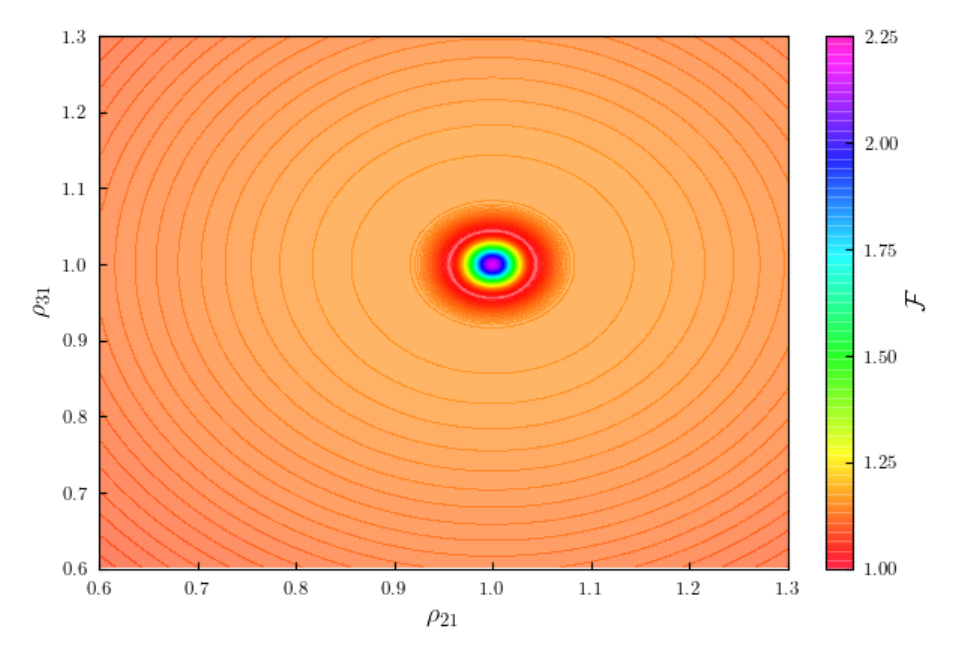}}\\
\caption{Representation of $\mathcal{F}$. (a) $\Lambda_{\rm s}/\Lambda_{\rm med}$ as a function of $\rho_{21}, \rho_{31}$. (b) 3D representation of the function $\mathcal{F,}$ which we use as an approximation for $\Lambda_{\rm s}/\Lambda_{\rm med}$. (c) Zoomed-in range of (b).}
\label{fig:fit_bns}
\end{figure*}

In SB2, we found that the estimates for the parameters of the compact binaries were biased. The chirp masses were overestimated, while the redshift was underestimated. The origin of this bias is described in the Appendix of SB2 (see Figure 13 of SB2). In order to find a fitting function, we proceeded as follows. We constructed a mock population of 1500 low-mass compact binaries according to the description given in Sec V of SB2. 

We chose to generate this mock population of low-mass compact binaries so as to have a wider range of S/Ns given the longer duration signals from these binaries. We chose the detection threshold such that the accumulated effective S/N $\rho_{\rm eff}>8,$ and the S/N for the $i^{th}$ segment in the $j^{th}$ detector $\rho^i_j> 3$ in at least one segment of 5 minutes in duration, for $j = (1,2,3)$ corresponding to the three ET detectors comprising the single ET. Then, for each segment, and for each of the detected binaries, we estimated $\frac{\Lambda_{\rm s}}{\Lambda_{\rm med}}$ (defined in eq. (\ref{lambda_s}) and eq. (\ref{lambda_med})) assuming that the measurement errors on the S/Ns are Gaussian, such that the standard deviations for $\rho$ is $\sigma_{\rho}=1$.

Figure \ref{fig:ratio_lambda} shows $\frac{\Lambda_{\rm s}}{\Lambda_{\rm med}}$ as a function of the S/Ns in the three ET detectors, $\rho_{21}$ and $\rho_{31}$ in the left and right panels, respectively. It can be clearly seen that for the bulk of the sources, $1 \lesssim \frac{\Lambda_{\rm s}}{\Lambda_{\rm med}} \lesssim 2$ for a very narrow range of $0.9 \lesssim \rho_{21} \lesssim 1.1$ and $0.9 \lesssim \rho_{31} \lesssim 1.1$. We therefore assume that $\frac{\Lambda_{\rm s}}{\Lambda_{\rm med}} = \mathcal{F}(\rho^i_{21}, \rho^i_{31})$, where $\mathcal{F}(\rho^i_{21}, \rho^i_{31})$ is as defined in eq. (\ref{fit_formula}).

In Figure \ref{fig:fit_3d} and \ref{fig:fit_2d}, we show the function $\mathcal{F}(\rho_{21}, \rho_{31})$ for all the segments of all the binaries that cross the detection threshold. Figure \ref{fig:fit_3d} shows the functional form, while Figure \ref{fig:fit_2d} shows the zoomed-in range to show the clear range of variation of $\mathcal{F}$ with $\rho_{21}, \rho_{31}$. It can be seen that including this function $\mathcal{F}(\rho^i_{21}, \rho^i_{31})$ in Equation 39 in SB2 as a prior information only affects a very narrow range where $0.95 \lesssim \rho_{21} \lesssim 1.05$ and $0.95 \lesssim \rho_{31}\lesssim 1.05,$ which is the bulk of population as seen in Figure \ref{fig:ratio_lambda}, and has negligible effect outside this range.

\section{Effect of $\mathcal{F}$ on mock populations }\label{sec:bias_effect}

\begin{figure*}[!]
\centering
\subfloat[\label{fig:ratio_mock1}]{\includegraphics[width=1.05\columnwidth]{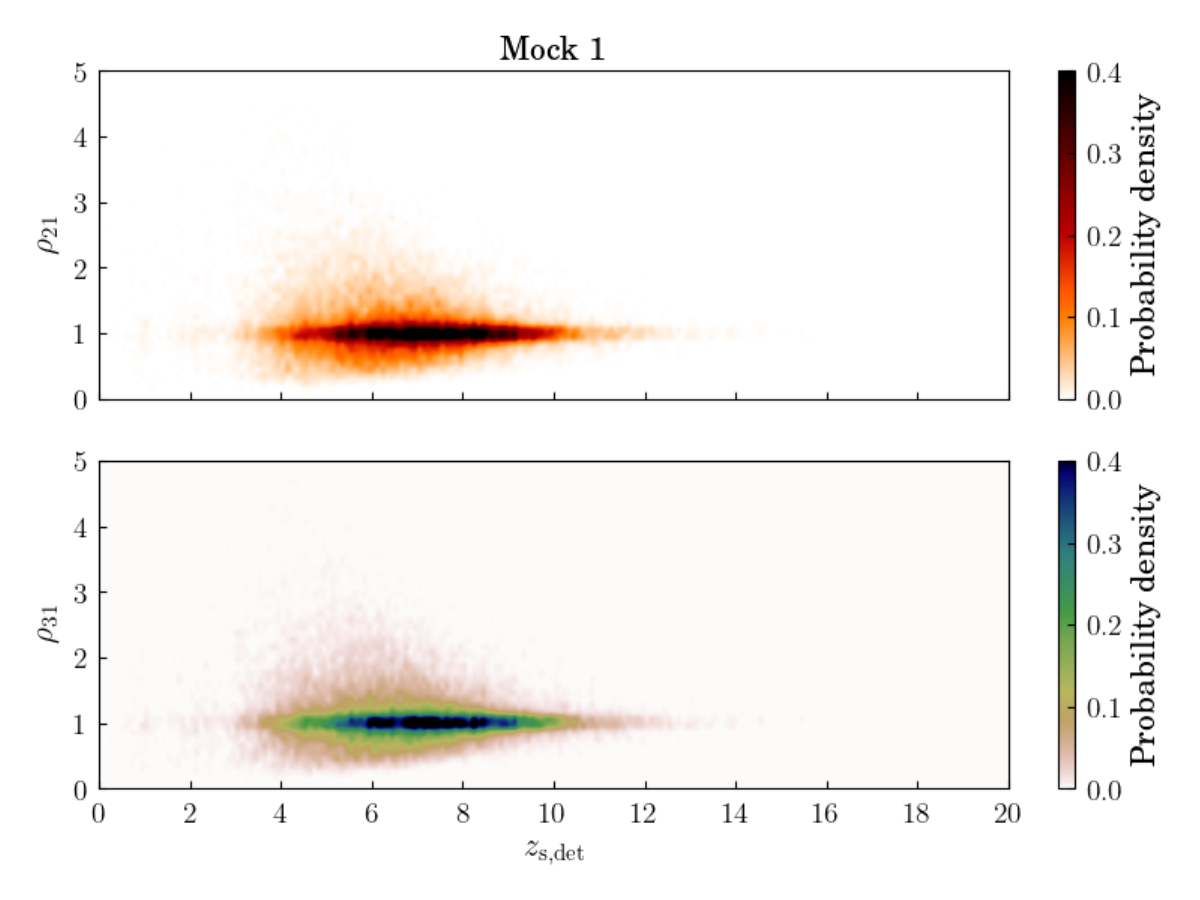}}\\
\subfloat[\label{fig:ratio_mock2}]{\includegraphics[width=1.05\columnwidth]{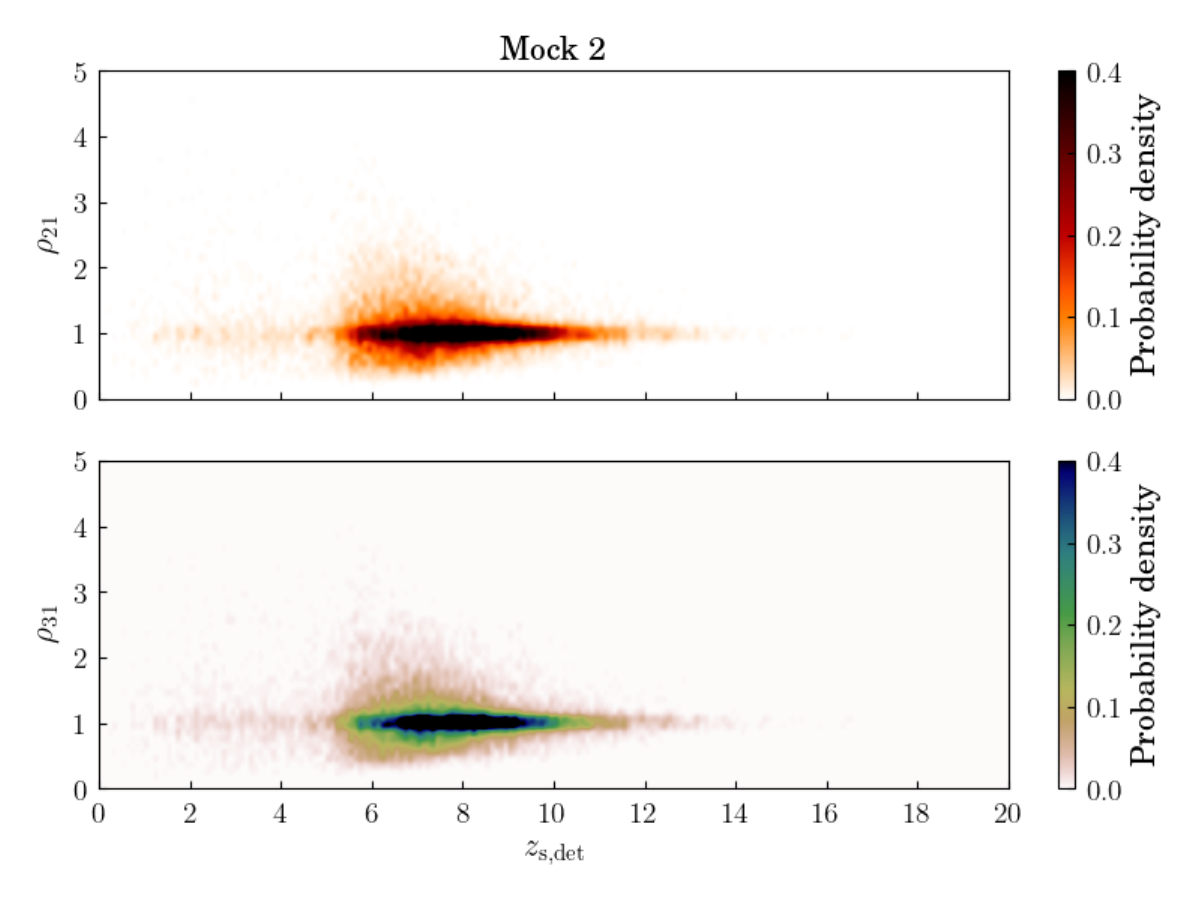}}\\
\subfloat[\label{fig:ratio_mock3}]{\includegraphics[width=1.05\columnwidth]{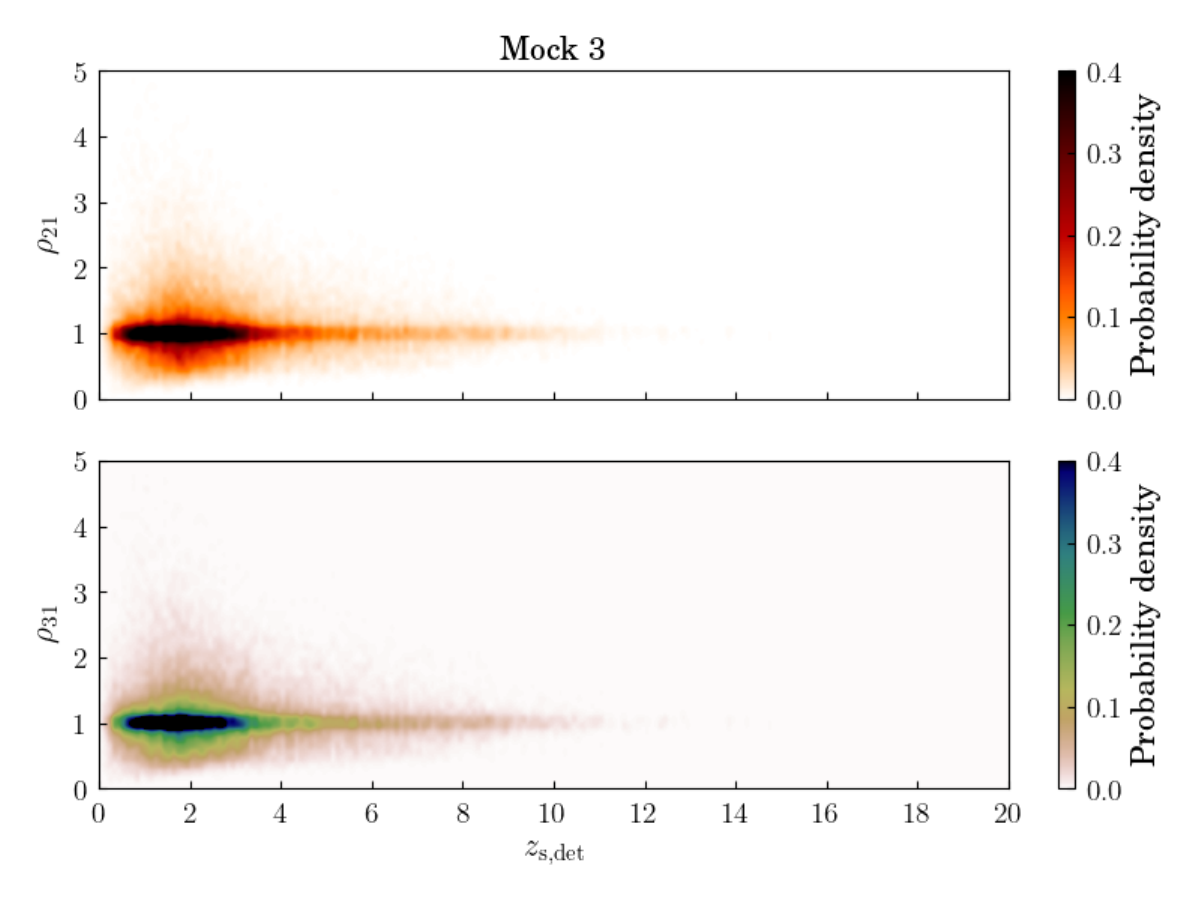}}\\

\caption{S/N as a function of redshift. Here, we show $\rho_{21}$ and $\rho_{31}$ as a function of redshift for compact binaries of (a) Mock 1, (b) Mock 2, and (c) Mock 3 where the subscript denotes one of the three ET detectors. }
\label{fig:ratio_redshift}
\end{figure*}

In this paper, we constructed multiple different mock populations as described in \S \ref{pop_description}. For each of these mock populations, we show the variation of $\rho_{21}$ and $\rho_{31}$ over the whole range of redshift in Figure \ref{fig:ratio_redshift}. As mentioned in section \S \ref{sec:bias_construct}, including $\mathcal{F}$ as prior information only affects a very narrow range where $0.95 \lesssim \rho_{21} \lesssim 1.05$ and $0.95 \lesssim \rho_{31}\lesssim 1.05$. As can be seen from Figure \ref{fig:ratio_redshift}, there are $\ll 1\%$ binaries at $z\sim 3$ for Mock 1, in the range $4 \lesssim z\sim \lesssim 6$ for Mock 2, and at $z\sim 13$ for Mock 3 and most of these generate S/Ns outside the range $0.95 \lesssim \rho_{ij} \lesssim 1.05$ where i,j are two of the  three ET detectors. 
Therefore, including $\mathcal{F}$ as prior information leads to no improvement in the bias for these redshift ranges in the respective mock populations.

\section{Reconstruction with extreme time-delay distributions}\label{extreme_delay}
\begin{figure*}
\centering
\subfloat[\label{fig:sfr_50_ideal_ex}]{\includegraphics[scale=0.45,right]{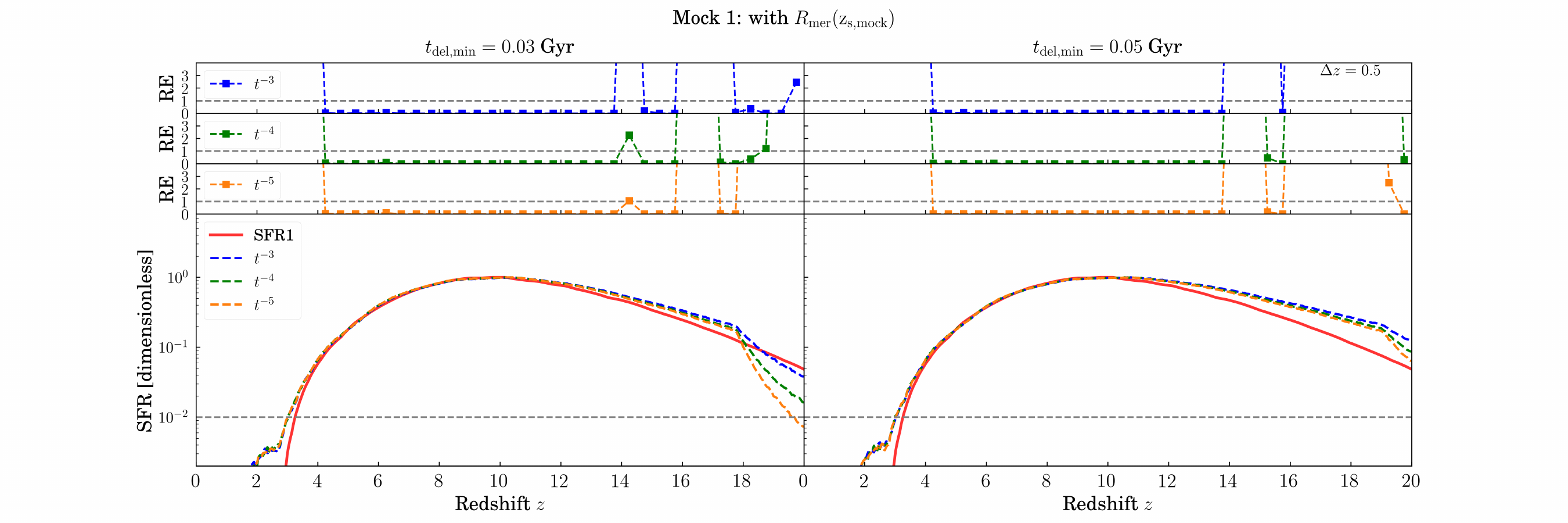}}\\
\subfloat[\label{fig:sfr_100_ideal_ex}]{\includegraphics[scale=0.45,right]{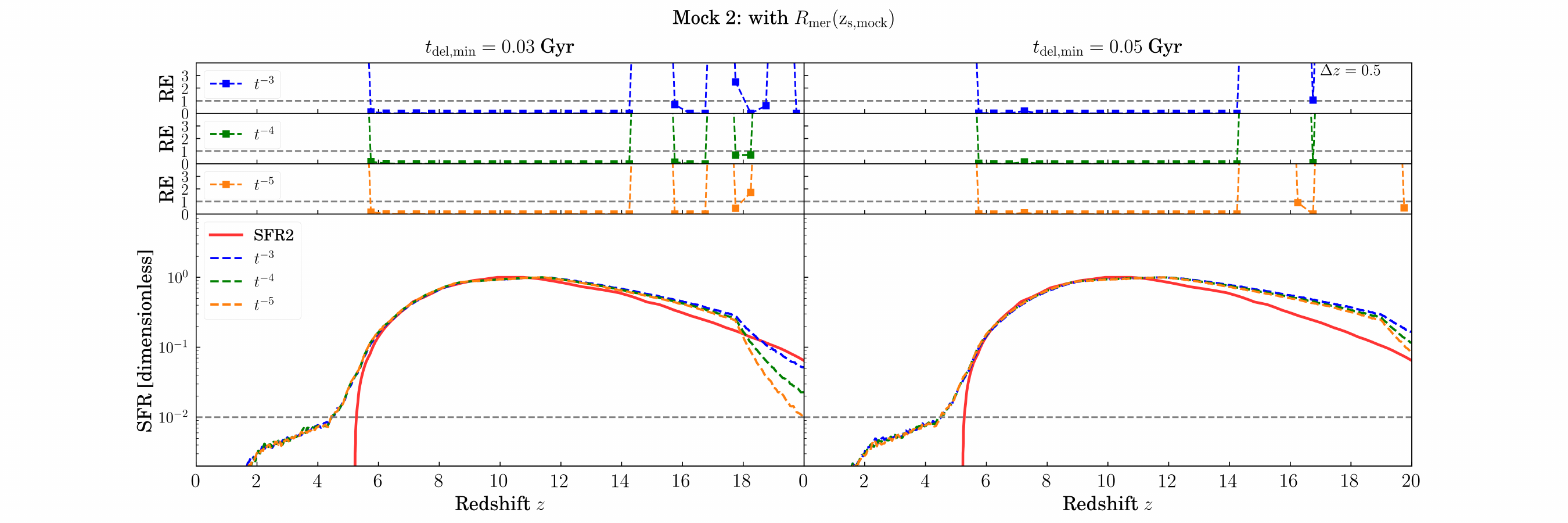}}\\
\subfloat[\label{fig:sfr_mixed_ideal_ex}]{\includegraphics[scale=0.45,right]{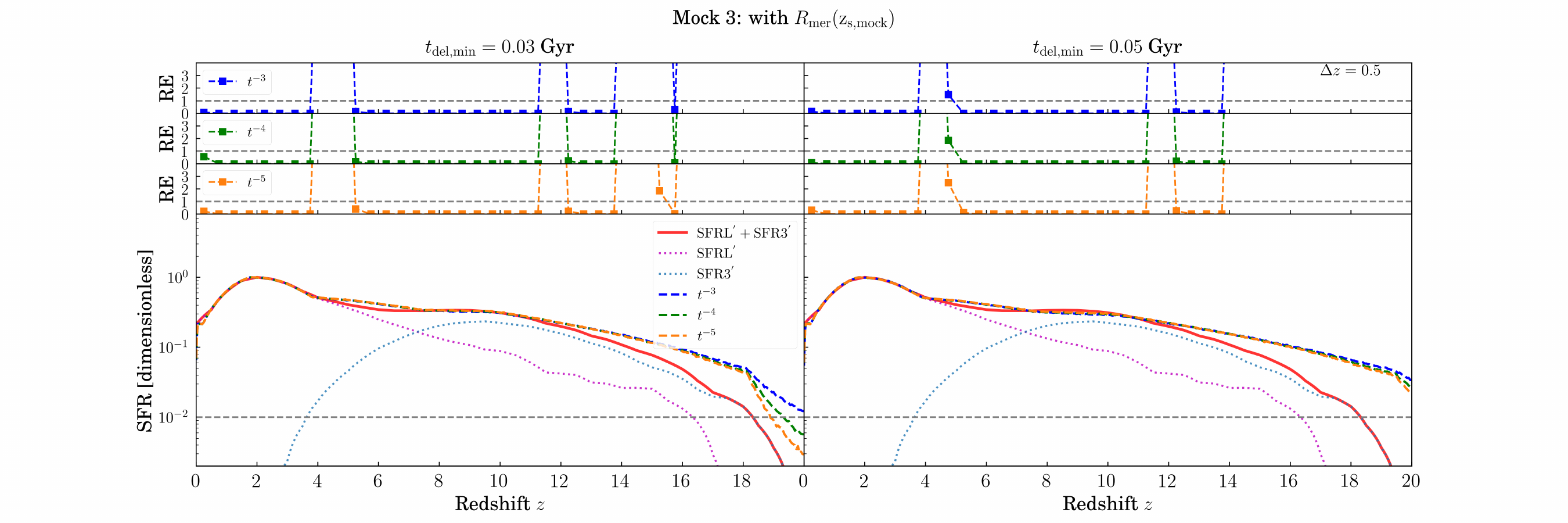}}\\
\caption{Reconstructed SFRs for the three populations assuming that the merger rate density is known with high accuracy. The maximum value of the SFR is normalised to 1. The left and right panels show the reconstruction assuming that $t_{\rm del, min} = 0.03$ and $0.05$ Gyr, respectively. The top three panels in all three plots show RE, or the KL divergence, for a redshift bin of  $\Delta z = 0.5$ in width. The red curve denotes the true SFR. We note that for the Mock 3 population, we estimate the merger-rate-weighted SFR.}
\label{fig:sfr_ex}
\end{figure*}

We concluded in \S \ref{sfr_rate}, based on Figure \ref{fig:sfr_ideal}, that our reconstruction of the SFR is essentially independent of the time-delay distributions up to $z \sim 14$ and that the accuracy of the reconstruction of SFR strongly depends on the time-delay distribution only at higher redshifts beyond $z\gtrsim 14$. In order to further prove this point, we assume three more time-delay distributions: (i) $t^{-3}$ , (ii) $t^{-4}$, and (iii) $t^{-5}$, where $t_{\rm del, min} < t < t_{\rm H}$ Gyr. The results for reconstructed SFR assuming these time delays and assuming that we know the merger rate as a function of redshift 
with high accuracy are shown in Figure \ref{fig:sfr_ex}. We can see that for these time-delay distributions, the accuracy of the reconstruction of SFR also strongly depends on the time-delay distribution only at higher redshifts of $z\gtrsim 14$.

\section{Acronyms used in the text}\label{sec:acronym_table}
Table \ref{tab:acronym} lists the various names of the variables used throughout the text in this paper.

\begin{table*}
\centering
\caption{List of variables used frequently in the text\label{tab:acronym}}
\def\arraystretch{1.5}
\begin{tabular}{|c|c|c|}
\hline
 Notation & Definition & Reference  \\
\hline

$t_{\rm ini}$ & cosmic time of the formation of a ZAMS binary & $t_{\rm ini} = t_{\rm obs} - t_{\rm del}$  \\
$t_{\rm obs}$ & cosmic time at which the compact binary is observed to merge.  &  \\
$t_{del}$ & delay time  &  $t_{\rm del} = t_{\rm evol} + t_{\rm merg}$  \\
$t_{\rm evol}$  & time of evolution from ZAMS to the formation of a compact binary system  &  \\
$t_{\rm merg}$ & time from formation of a compact binary till the merger  &   \\
$\mathcal{R}(z)$  & merger rate density per unit redshift as a function of redshift  &  Eq. \ref{Rofi}  \\
${\rm{SFR}}(z)$  & star formation rate density as a function of redshift  &  Figure \ref{fig:sfr_all} \\
Z  & metallicity &  \\
$M_{\rm sim}$  & total mass of all stars accompanying the stellar evolution, & \\
& leading to formation of compact object binaries  &  \\
& including the binaries as well as the single stars. &  \\
$\mathcal{M}_{\rm s,mock}$, $M_{\rm s, mock}$, $z_{\rm s, mock}$ & subscript (${\rm s,mock}$) denotes true source parameters &   \\
&  of a binary in the mock population &   \\
$\mathcal{M}_{\rm s,det}$, $M_{\rm s,det}$, $z_{\rm s, det}$  &  subscript (${\rm s,det}$) denotes true source parameters  &   \\
 &  of detected sources &   \\
$\mathcal{M}_{\rm med,det}$, $M_{\rm med,det}$, $z_{\rm med, det}$  &   subscript (${\rm med,det}$) denotes the median of the estimated  &  see Fig. 5 in SB2  \\
&  posterior distribution for each parameter  & for an example of \\

&  & posterior distributions \\
$N_{\rm mock}$ & number of binaries in a mock population  &   \\
$N_{\rm det}$  & number of detected binaries &   \\
$N_{\rm yr}$  &  number of mergers per year &   \\

$T_{\rm mock}$  & time during which binaries in the mock population are expected to merger  &  Eq. 31 \\
$R_{\rm mer}(z_{\rm s, mock})$ &  merger rate densities as a function of redshift  &  Eq. \ref{merger_rate_den} \\
 & assuming true source redshift in the mock population &   \\
$R_{\rm mer}(z_{\rm s, det})$  & merger rate densities as a function of redshift  & Eq. \ref{merger_rate_den}  \\
 & assuming true sources redshift of detected sources &   \\
$R_{\rm mer}(z_{\rm med, det})$ & merger rate densities as a function of redshift with the median & Eq. \ref{merger_rate_den}  \\
 & of the estimated posterior distribution of the redshift of detected sources &   \\
subscript 'sec' & parameters of the secondary population &   \\
$\mathcal{D}(z_i, z_{i+1})$ & detection efficiency in the redshift bin ($z_i, z_{i+1}$)  &  Eq. \ref{det_eff} \\ 
$R_{\rm mer, recon}$ &  reconstructed merger rate density taking into account the detection efficiency &  Eq. \ref{recons_mer} \\
$\rm SFR^{'}$ &  merger-rate-weighted SFR &  \\

\hline
\hline
\end{tabular}
\end{table*}

\end{appendix}

\end{document}